\documentclass[acmsmall]{acmart}
\usepackage{cite}
\usepackage{amsmath,amssymb,amsfonts}
\usepackage{algorithmic}
\usepackage{graphicx}
\usepackage{subfigure}
\usepackage{textcomp}
\usepackage{xcolor}
\usepackage{bm}
\usepackage{amsmath}

\usepackage{algorithm}
\usepackage{url}
\usepackage{breakurl}
\usepackage{booktabs}
\usepackage{xspace}
\usepackage{balance}
\usepackage{multirow}
\usepackage{colortbl}

\newcommand{\revised}[1]{\textcolor{black}{#1}}

\newcommand{\mytool}{\textsc{XMal}\xspace}
\usepackage{listings}
\definecolor{codegreen}{rgb}{0,0.6,0}
\definecolor{codegray}{rgb}{0.5,0.5,0.5}
\definecolor{codepurple}{rgb}{0.58,0,0.82}
\definecolor{backcolour}{rgb}{0.95,0.95,0.92}
\usepackage{textcmds}
 
\lstdefinestyle{mystyle}{
commentstyle=\color{codegreen},
keywordstyle=\color{magenta},
numberstyle=\tiny\color{codegray},
stringstyle=\color{codepurple},
basicstyle=\ttfamily\footnotesize,
breakatwhitespace=false,         
breaklines=true,                 
captionpos=b,                    
keepspaces=true,                 
numbers=left,                    
numbersep=5pt,                  
showspaces=false,                
showstringspaces=false,
showtabs=false,                  
tabsize=2,
frame=single,
frameround=fttt,
xleftmargin=10pt,
xrightmargin=10pt
}

\usepackage{hyperref}
\hypersetup{
colorlinks=true,
linkcolor=blue,
urlcolor=blue,
citecolor=blue
}

\setcopyright{none}

\begin{document}
\title{Why an Android App is Classified as Malware? Towards Malware Classification Interpretation}

\author{Bozhi Wu}
\email{wubozhi@pku.edu.cn}
\affiliation{%
  \institution{Nanyang Technological University, Singapore and Peking University, China}
}

\author{Sen Chen}
\email{chensen@ntu.edu.sg}
\affiliation{%
  \institution{College of Intelligence and Computing, Tianjin University, China and Nanyang Technological University, Singapore}
}
\authornote{Sen Chen is the corresponding author.}


\author{Cuiyun Gao}
\email{cygao@cse.cuhk.edu.hk}
\affiliation{%
  \institution{Harbin Institute of Technology (Shenzhen)}
  \city{Shenzhen}
  \country{China}
}

\author{Lingling Fan}
\email{llfan@ntu.edu.sg}
\affiliation{%
  \institution{College of Cyber Science, Nankai University, China and Nanyang Technological University, Singapore}
}

\author{Yang Liu}
\email{yangliu@ntu.edu.sg}
\affiliation{%
  \institution{Nanyang Technological University}
  \country{Singapore}
}

\author{Weiping Wen}
\email{weipingwen@pku.edu.cn}
\affiliation{%
  \institution{Peking University}
  \country{China}
}

\author{Michael R. Lyu}
\email{lyu@cse.cuhk.edu.hk}
\affiliation{%
  \institution{Chinese University of Hong Kong}
  \city{Hong Kong}
  \country{China}
}


\renewcommand{\shortauthors}{Wu and Chen et al.}

\begin{abstract}
Machine learning (ML) based approach is considered as one of the most promising techniques for Android malware detection and has achieved high accuracy by leveraging commonly-used features. In practice, most of the ML classifications only provide a binary label to mobile users and app security analysts. However, stakeholders are more interested in the reason why apps are classified as malicious in both academia and industry. This belongs to the research area of interpretable ML but in a specific research domain (i.e., mobile malware detection). Although several interpretable ML methods have been exhibited to explain the final classification results in many cutting-edge Artificial Intelligent (AI) based research fields, till now, there is no study interpreting why an app is classified as malware or unveiling the domain-specific challenges.

In this paper, to fill this gap, we propose a novel and interpretable ML-based approach (named \mytool) to classify malware with {high} 
accuracy and explain the classification result meanwhile. (1) The first classification phase of \mytool hinges multi-layer perceptron (MLP) and attention mechanism, and also pinpoints the key features most related to the classification result. (2) The second interpreting phase aims at automatically producing neural language descriptions to interpret the core malicious behaviors within apps. \revised{We evaluate the behavior description results by leveraging a human study and an in-depth quantitative analysis. Moreover, we further compare \mytool with the existing interpretable ML-based methods (i.e., Drebin and LIME) to demonstrate the effectiveness of \mytool}. 
{We find that \mytool is able to reveal the malicious behaviors more accurately.} 
Additionally, our experiments show that \mytool can also interpret the reason why some samples are misclassified by ML classifiers. Our study peeks into the interpretable ML through the research of Android malware detection and analysis.
\end{abstract}


\maketitle

\section{introduction}
Android malicious applications (malware) have become a serious security issue as the mobile platform has become increasingly popular~\citep{ref_url1}. For example, more and more app users store personal data such as banking transactions on their mobile devices~\citep{chen2019ausera, chen2018mobile}, consequently, hackers shift their attention on mobile devices and try to perform malicious behaviors through Android apps. It is not surprising that a number of approaches have been proposed for detecting Android malware. Specifically, traditional signature-based approaches~\citep{schlegel2011soundcomber, zhou2012hey, zhou2013fast} require frequent updates of the signature database and fail to be effective in detecting emerging malware. Behavior-based approaches~\citep{yan2012droidscope, wu2014airbag, tam2015copperdroid, graziano2015needles} also rely on the predefined malicious behaviors, which is limited by the analysis of existing malicious samples. Data flow-based approaches~\citep{arzt2014flowdroid, li2015iccta, gordon2015information} are usually used to identify data leakage related malicious behaviors. Recently, researchers have proposed many effective Android malware detection methods by using a plethora of machine learning (ML) algorithms (e.g., KNN~\citep{aafer2013droidapiminer}, SVM~\citep{arp2014drebin}, Random Forest~\citep{rastogi2013droidchameleon}, and XGboost~\citep{fereidooni2016anastasia}) to classify and categorize malware. In these approaches, Android permissions and API calls are the commonly-used feature types~\citep{wu2012droidmat, arp2014drebin, chen2016stormdroid, chen2018automated}, and achieved a high detection accuracy (more than 90\%). Meanwhile, researchers began to leverage deep neural networks like CNN and RNN (e.g., LSTM and GRU) to detect Android malware~\citep{kim2018multimodal, yuan2016droiddetector,fengmobidroid,feng2020seq,feng2020mobitive} and 
promising performance {has been achieved}.

However, these ML-based methods only provide a binary label to mobile users and app security analysts. In other words, these existing methods do not completely solve the problem of malware detection because they merely mean that the classified apps are most likely Android malware or benign apps. In practice, in many cases, only knowing the classification results is not enough. For example, (1) the app store needs to know exactly what malicious behaviors the apps employ, instead of classification results, in order to decide whether to remove them from markets. (2) For app security analysts, they need to identify various malware and then understand the malicious behaviors manually with substantial effort. It is a difficult and time-consuming task to analyze a large-scale dataset of Android malware in the wild. However, the truth 
is that millions of malware are classified and stored in the server.
Therefore, interpreting and understanding what an ML model has learned and how the model makes prediction can be as important as the detection accuracy since it can guarantee the {reliability} of the classification model. 
Additionally, the robustness of ML models is facing the security threat of adversarial samples according to a large number of relevant research including Android malware~\citep{chen2016towards, chen2018automated, chen2019can, hu2017generating,chen2019real,lei2020advanced}.
As ML-based methods are black-box and cannot explain how they make predictions, adversaries might fool these methods by constructing a little perturbation to misclassify malware as benign samples more smoothly.

In order to solve the problems mentioned above, we first investigated the approach of interpreting malicious behaviors in Drebin \citep{arp2014drebin} and found that the approach localized malicious behavior from the trained model rather than the test sample itself. 
After that, we tried to explain the classification of the malware detection using an interpretable ML method called LIME~\citep{ribeiro2016should}, but the feature results are mismatched to the behavior because LIME did not consider the correlation between the input features. 
To take the correlation between different features into account, we find that attention mechanism has been applied in machine translation and computer vision (CV), and achieved great success of interpretability~\citep{arras2017relevant, zhou2018interpretable, DBLP:conf/ijcnlp/GhaderM17, DBLP:conf/icml/XuBKCCSZB15}. 
Therefore, we follow this research line, and propose a novel and interpretable ML-based approach (named \mytool) to detect Android malware and interpret how predictions are made. {\mytool leverages a customized attention mechanism with a multi-layer perceptron (MLP) model, which pinpoints the key features most related to the prediction result, since the traditional attention mechanism cannot be used directly in Android malware detection scenario (\S \ref{subsec:motivation}). } 
Apart from the binary result, it also automatically generates a 
{descriptive}
explanation (i.e., a malicious behavior description) for the classification according to the key features. 
Additionally, it can help to explain why some benign apps are misclassified as malware and vice versa. We conduct comprehensive experiments to demonstrate its interpretability of Android malware detection, and the results show that \mytool can detect Android malware effectively, with 98.35\% accuracy, and can identify the malicious behaviors that are validated by 
\revised{a human study through an online survey. Our quantitative analysis can also be used to demonstrate the better performance on malware description generation of \mytool.}
In addition, we compare the results with the state-of-the-art {techniques} in the interpretability of Android malware detection scenario. Finally, we present case studies and in-depth discussion {about}
our approach.

In summary, we make the main contributions as follows.

\begin{itemize}
    \item We are the first work focusing on the interpretability of Android malware detection and analysis. We concentrate on why an Android app is classified as malware rather than 
    the detection accuracy {only}.
    
    \item {We propose \mytool to interpret the malicious behaviors of Android malware, by leveraging a customised attention mechanism with multi-layer perceptron (MLP).}
    
    \item \revised{We conduct a human study by designing an online survey and a quantitative analysis to validate the capability of \mytool regarding interpretability, and also provide an in-depth comparison study with the state-of-the-art techniques to demonstrate the effectiveness of \mytool.}
    
    \item We present several case studies and an in-depth discussion to highlight the lessons learned and the current status of interpretability of Android malware detection and analysis.
    
\end{itemize}

\section{Background}
In this section, firstly we review several potential solutions for interpretability in Android malware detection and point out their weaknesses. Secondly, we introduce the attention mechanism as our work uses the concept of attention mechanism. Finally, we highlight the motivation of our work. 

\subsection{Potential Solutions for Interpretability in Android Malware Detection}
ML technique is widely used to classify the samples into different categories, however without explaining the reason for the prediction results (i.e., not \textit{interpretable}). \textit{interpretable}, defined by Doshi-Velez et al.~\citep{doshi2017towards}, is the ability to explain or present the results in understandable terms to human. {In order to alleviate this problem, some general methods which are model-agnostic have been proposed, such as LIME \citep{ribeiro2016should} and LEMNA \citep{guo2018lemna}.
On the other head, researchers have also done some studies in {areas of} text categorization and image classification.} For example, Arras et al.~\citep{arras2017relevant} tried to demonstrate that understanding text categorization can be achieved by tracing the classification decision back to individual words using layer-wise relevance propagation (LRP), a recently developed technique for explaining predictions of complex non-linear classifiers. Zhou et al.~\citep{zhou2018interpretable} proposed a new framework called Interpretable Basis Decomposition for providing visual explanations for image classification networks. By decomposing the input image into semantically interpretable components, the proposed framework can quantify the contribution of each component to the final prediction.

In Android malware detection and analysis, malware is identified by features (e.g., permissions, intents, and API calls) extracted from the APK file.
Usually, app analysts first extract dangerous permissions and intents from AndroidManifest.xml. They utilize existing tools (e.g., \textsc{dex2jar}) to decompile Dalvik executable (dex) files in the Android application package (apk) file to get the source code and read the source code from the beginning to end to locate malicious code segments that lead to malicious behaviors. Finally, they can identify malware through malicious behaviors, which is very understandable to a human. In order to explain the predictions in ML, some key permissions, APIs, intents, or code segments should be used to match certain behaviors of Android apps, which help us understand what behaviors the Android app might perform, causing it to be classified as malware. Therefore, to explain why an app is classified as malware, we need to find out which features have a significant impact on the classification in ML, and whether they are indeed related to malicious behaviors of the malware. In order to do that, Drebin~\citep{arp2014drebin} utilized the simple detection function of linear SVM to determine the contribution of each individual feature to the classification result, which can be used to explain the classification of Android malware. However, since Drebin actually outputs the features with the highest weights in the ML classifier, rather than the test samples, the feature weights of different test samples are the same, which may be inaccurate.
Melis et al.~\citep{melis2018explaining} proposed to leverage a gradient-based approach to identify the most influential local features. This method essentially obtains the gradient by approximating the original complex model, and there is inevitably a bias. In summary, there is no specific study on the interpretability of Android malware detection and analysis to interpret their corresponding malicious behaviors so far.

\subsection{Attention Mechanism}
Attention mechanism is a fairly popular concept and useful tool in the DL community in recent years \citep{ref_url2}.
In deep learning (DL), it refers to paying more attention to certain factors when processing data. It utilizes the attention vector to estimate how much an element is related to the target or other elements, and take the sum of their values weighted by the attention vector as the approximation of the target.

\begin{figure}
\centering
\includegraphics[width=0.5\textwidth]{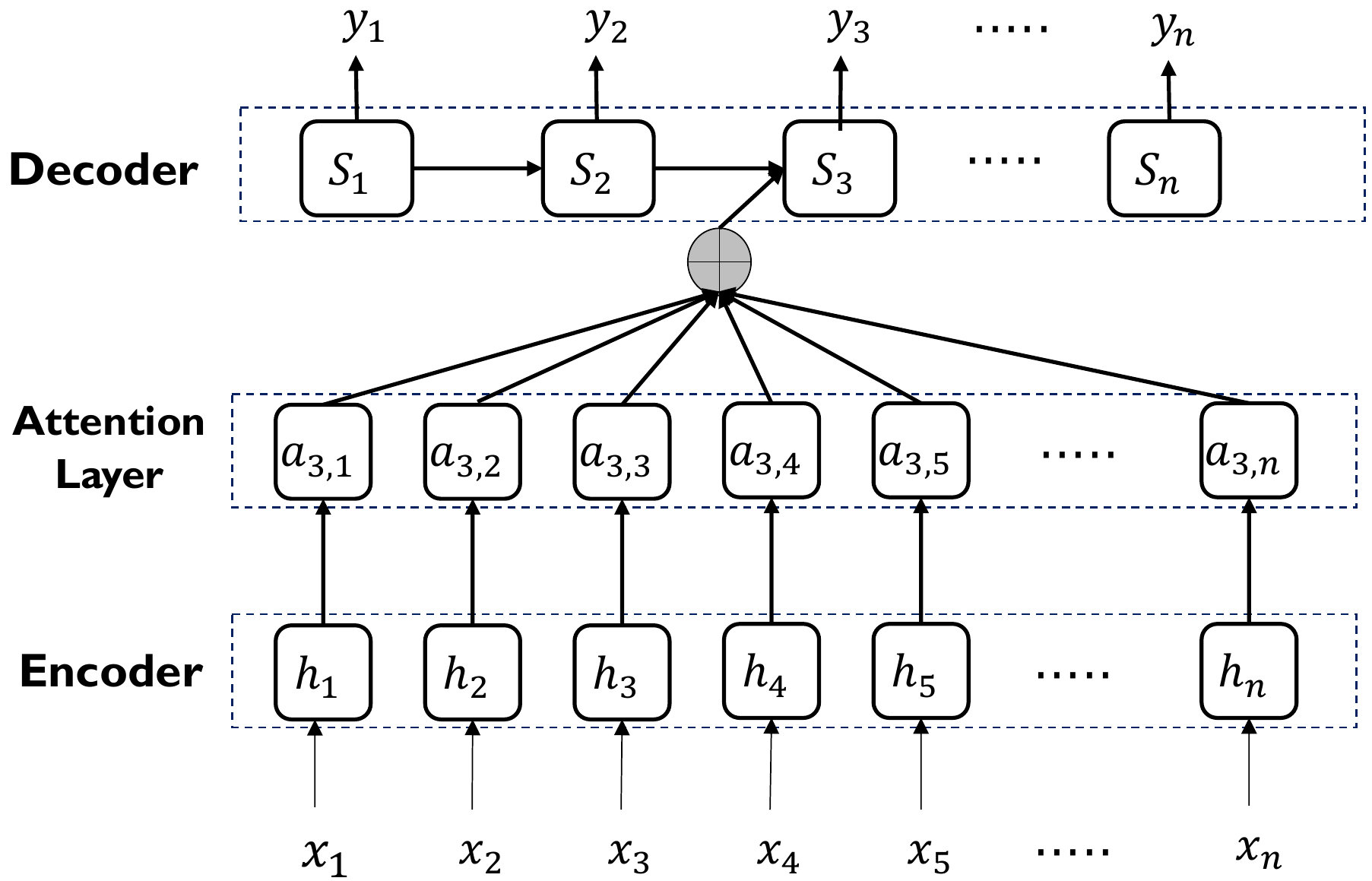}
\caption{Attention in machine translation}
\label{fig2:attention}
\end{figure}

It was first proposed by Bahdanau et al. \citep{bahdanau2014neural} to solve the problem of incapability of remembering long source sentences in neural machine translation (NMT). An attention layer is embedded between the encoder layer and the decoder layer, as shown in Fig.~\ref{fig2:attention}.
The attention vector $c_i= \{a_{i,1}, a_{i,2}, a_{i,3}, a_{i,4},.., a_{i,n} \}$ has access to the entire input sequence, which guarantees the ability of remembering long source sentences. More importantly, it also shows how significantly an input element is related to the output target, and which input element is more important or has a higher weight to generate the output.

{Attention mechanism shows superiority in terms of classification and interpretability. It can help the model assign different weights to each part of the input, extract more critical and important information, making the model's predictions more accurate, and make the prediction more understandable. For example, Xu et al. \citep{xu2015show} proposed a method to explain why a certain word is output by visualizing the attention weights of the image region.
This is why the attention mechanism is so popular. In this paper, we {make the} first attempt to use and customize attention mechanism in Android malware detection and analysis in order to interpret the prediction results.}

\subsection{Motivation of Our Work}\label{subsec:motivation}
In order to interpret the malware classification results, most existing interpretable ML-based methods utilize linear models or simple models (e.g., decision trees and linear regression)
to approximate the original complex model~\citep{ribeiro2016should}, because these models can simply show the weight of each feature that contributes to the classification results. However, the usage of these models to approximate the original complex model inevitably introduces deviations.
Additionally, most of these methods do not take into account the correlation between the input features. In fact, the features used by Android malware detection are usually highly correlated such as {SmsManager.sendTextMessage} and {android.permission.SEND\_SMS}. This leads to the inability of these methods to give a correct explanation for Android malware detection. In order to address these problems and challenges, we propose a novel and effective method by using the attention mechanism with MLP for Android malware detection.
{The attention mechanism estimates how strongly a feature is correlated with other features and how important a feature is related to the prediction result. In Android malware detection scenario, we try to customize the attention mechanism through a fully connected network to learn the correlation between scalar-valued elements and assign corresponding weights to elements, since the traditional attention mechanism is performed on elements in the form of vectors and cannot be used directly in this case. }

\section{Approach}
{In this section, we first introduce the overview of our approach (named \mytool), 
and then the details of each component. }

\begin{figure*}
\centering
\includegraphics[width=0.85\textwidth]{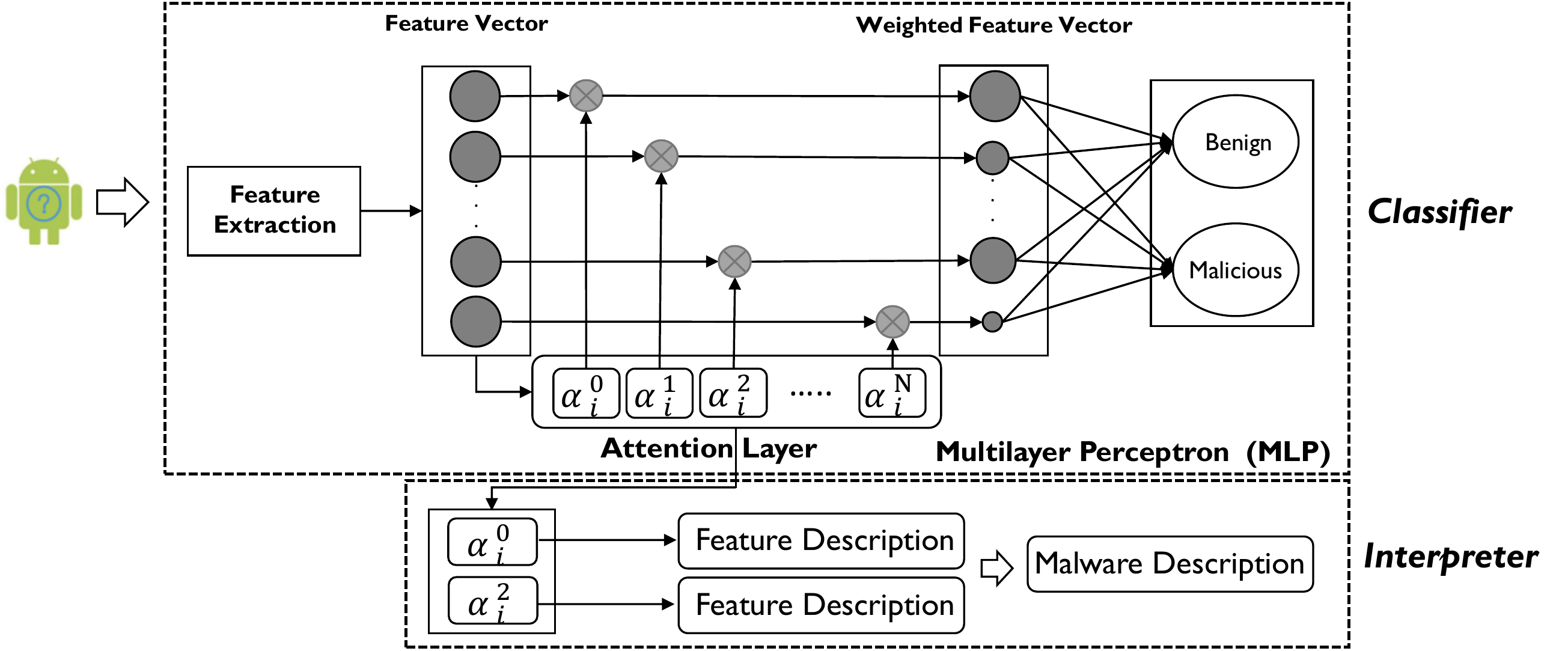}
\caption{Overview of our approach (\mytool)}
\label{fig:overview}
\end{figure*}

\subsection{Overview}
As shown in Fig.~\ref{fig:overview}, our approach (\mytool) consists of two main components (i.e., a {Classifier} and an {Interpreter}). (1) The classifier component extracts API calls and used permissions from APK files as inputs, and aims at accurately predicting whether an app is malware. The classifier can also pinpoint the key input features most related to the prediction result. (2) The interpreter component aims at automatically producing descriptions to interpret why an app is classified as malware. The behavior descriptions are generated through the {rule-based method} according to the documentation collected from Android Developers~\citep{ref_url3}. The details of each component are elaborated in Section~\ref{ssec:classifier} and Section~\ref{ssec:description}, respectively.

\subsection{Classifier Component}\label{ssec:classifier}
In this section, we introduce how to extract the key features that have more relevance to the classification results.
The key feature extraction conducts two processes: feature extraction and model training. We detail the two processes as below.

\subsubsection{Feature Extraction}
Usually, if an Android app exhibits malicious behaviors, it will be granted with the necessary permissions and call the corresponding APIs.
In fact, permission and API calls are the top two important and commonly-used feature types for Android malware detection and analysis \citep{peiravian2013machine}. A lot of studies used these two features as significant features for classifying Android malware, such as Drebin~\citep{arp2014drebin}, DriodAPIMiner~\citep{aafer2013droidapiminer}, DroidMat~\citep{wu2012droidmat} and many other previous studies~\citep{fereidooni2016anastasia,kim2018multimodal,yuan2016droiddetector, chen2018automated, yerima2013new, chen2016stormdroid,mclaughlin2017deep,xu2018deeprefiner}.
Additionally, they contain semantics that can be used to help to understand the behaviors of the application. 
Therefore, in this paper, we follow the common practice and use API calls and permissions as the features to train a malware classifier.
In Android system, there are hundreds of permissions,
and the number of APIs exceeds 20,000.
But not all of them are helpful in distinguishing malware. Li et al. \citep{li2018significant} utilized three levels of pruning and found that only 22 permissions are significant for detecting malware. Therefore, we need to employ pruning to preserve those features that can be used to identify malware efficiently. 
Here we refer to the paper \citep{chen2018automated} and select 158 features (including 97 API calls and 61 permissions) for our study {by using manual statistical pruning method in \citep{chen2018automated} from the original 2,114 features extracted from the training sample set.}
The selected features have a high degree of discrimination for malware classification, which is good for improving the accuracy and interpretability of the classification. 
\revised{Meanwhile, since API calls and permissions have more semantics that enable people to understand their role in the applications, using them as the features can help further interpret our model.} 
\revised{Additionally, our approach is general that can be extended the new feature categories to capture more complex malicious behaviors according to new malware samples.}
In order to extract API calls and permissions, we utilize {Androguard}~\citep{desnos2013androguard} to extract API calls and permissions from APK file, which are used to construct the feature vector. 
Here we denote a sample set by 
\revised{$\{\{(\bm{x_i},y_i)\}|\bm{x_i}\in X,y_i\in Y, 1<=i<=M\}$, where $X$ is the set of $\bm{x_i}$ and $Y$ is the set of $y_i$}, $ \bm{x_i} = (x_i^{(1)},x_i^{(2)},x_i^{(3)}, ... , x_i^{(N)}) $ is the feature vector of the $ i$-th sample, $ N $ is the total number of features, $ y_i \in \{0,1\} $ is the label of the $ i$-th sample (i.e., 0 for benign, 1 for malicious), and $ M $ is the total number of samples. $ x_i^{(j)} $ represents the $ j$-th feature of  the $ i$-th sample. If the $ j$-th feature exists in the $ i$-th sample, then $ x_{i}^{(j)} = 1 $, otherwise, $ x_{i}^{(j)} = 0 $.

\subsubsection{Customized Classification Model and Model Training}
{After extracting features and constructing a feature vector, we feed the feature vector to train the malware classifier. As shown in Fig. 2, the classifier consists of two layers: the \emph{attention layer} and the \emph{multi-layer perceptron (MLP)}. The attention layer is designed to learn weights of the features which can be regarded as relevancy scores between the features and classification results. Then the MLP maps the features weighted by the attention layer to the binary classification.}

{The traditional attention mechanism is to obtain the weight of the input feature by scoring how well the input feature and the output match, which can be formulated as follows:
\begin{equation}
\bm{e_{ij}}= score(\bm{s}_{i-1}, \bm{h_j}),
\end{equation}}

{where}
$\bm{s}_{i-1}$ is hidden state of output, and $\bm{h_j}$ is the j-th annotation of input. Then the feature weight can be computed by:
\begin{equation}
    \alpha_{ij} = \frac{exp(e_{ij})}{\sum_{k=1}^{n}{exp(e_{ik})}}.
\end{equation}
{The score function will be different according to different scenarios. For instance, the score function in the paper {by Luon et al.} \citep{luong2015effective} is computed by:
\begin{equation}\label{eq2}
score(\bm{s}_{i-1}, \bm{h_j}) = \bm{s}_{i-1}^T \bm{h_j},
\end{equation}}
{The input feature of traditional attention mechanism is generally expressed as a vector. But the features extracted from the samples are composed of scalar values. They can not be used to compute the score like Equation ~\ref{eq2}. Here we customize a fully connected {network} and a softmax function to implement the attention layer, as shown in Fig.~\ref{fig:attention_layer}. {Because a fully connected network can capture the correlations between scalar-valued input features.}

\begin{figure}
\centering
\includegraphics[width=0.5\textwidth]{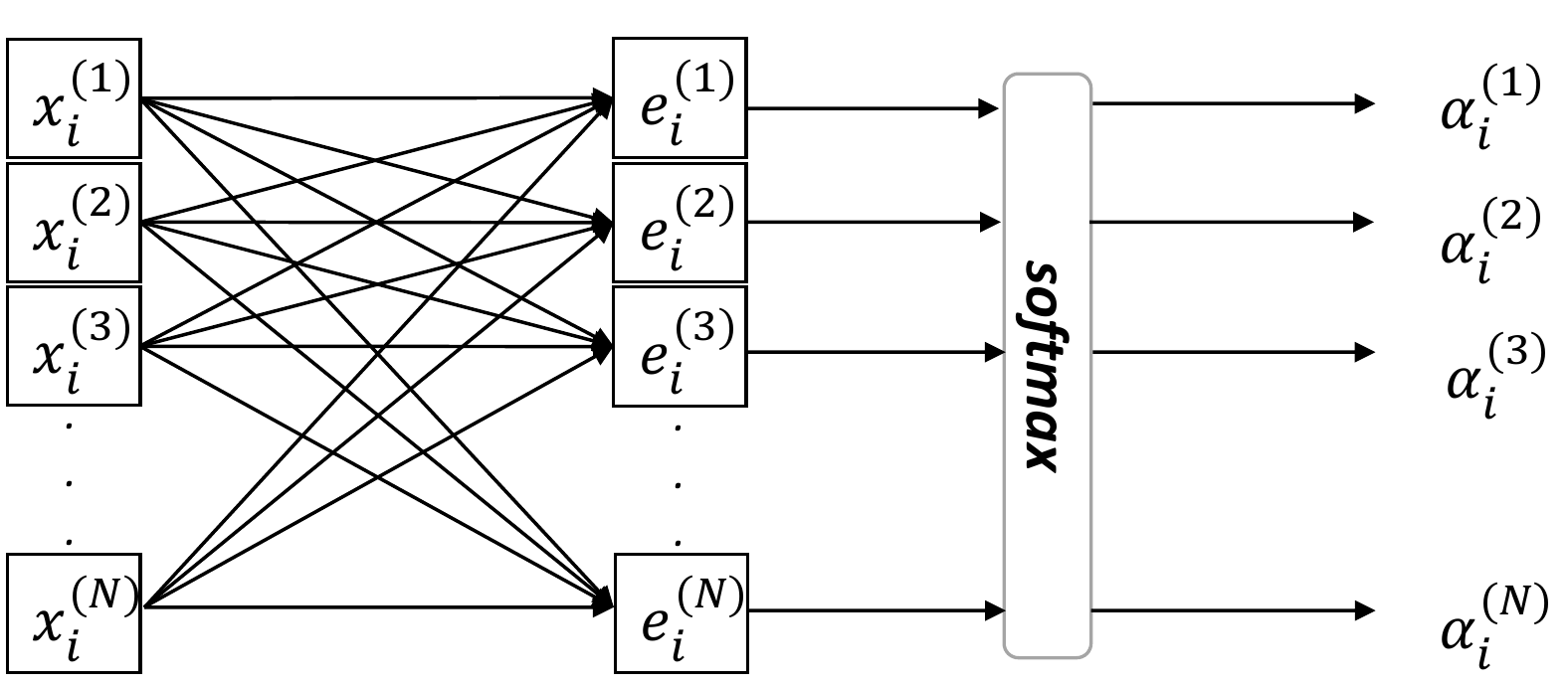}
\caption{Attention layer in \mytool}
\label{fig:attention_layer}
\end{figure}}

{We compute how well all input features and the output at j-th position match by:
\begin{equation}\label{eq3}
e_i^{(j)}=\ \sum_{k=1}^{N}x_i^{(k)}w_{kj},
\end{equation}}
{where}
$ w_{kj} $ is a learnable parameters of the fully connected {network} in attention layer. $e_i^{(j)}$ as the output at j-th position in the fully connected {network}, is a linear combination of all input features $x_i^{(k)}$. It can be regarded as the combination of a set of features that have different relevance to the input feature at j-th position. After the model training, the parameter $w_{kj}$ will be assigned an appropriate value to show the correlation between the input feature at j-th position and other input features. Therefore, our customized attention layer has considered the correlation between the input features when computing the weight of input features.

{Here we perform a softmax function on the output of the fully connected {network} to obtain the weights of input features at different positions. We denote attention vector by $\bm{\alpha_i}$, where $\bm{\alpha_i}= (\alpha_i^{(1)}, \alpha_i^{(2)},\alpha_i^{(1)}, ... , \alpha_i^{(n)})$. $\alpha_i^{(j)}$ represents the weight of $ j$-th feature in $ i$-th sample, and is computed by:
\begin{equation}
 \alpha_i^{(j)}= \frac{exp(e_i^{(j)})}{\sum_{k=1}^{N}{exp(e_i^{(k)})}},   
\end{equation}}
    {where}
$\alpha_i^{(j)}$ reflects the importance of the input feature at i-th position in deciding classification results.

{After generating the attention vector through the attention layer, the MLP is used to map the features weighted by the attention vector to the binary classification. Here we denote the weighted feature vector of the i-th sample by $\bm{c_i}$. It is obtained by weighting the input feature vector {using}
the attention vector, and is computed by:
\begin{equation}
    \bm{c_i}=\bm{\alpha_i} \bm{x_i}^T.
\end{equation}
}

{In the end, the classification result can be computed by:
\begin{equation}
     \bm{y}_i=\emph{f }(\bm{c_i}),
\end{equation}
}
{where $\emph{f }(\cdot)$ represents the function of MLP that maps the input vector $\bm{c_i}$ into a binary prediction result.}

When the training data are fed to train the classifier, the attention layer assigns different weights to the corresponding features based on their relevance to the classification result. Features that have more relevance to classification are assigned {larger weights}, while features with less impact are assigned {smaller weights}. Other interpretable ML methods aim to obtain the weight of the feature 
by approximating the original complex model. Unlike them, \mytool directly obtains the weight of the feature by embedding the attention layer in the model, therefore there is no deviation.
After feature extraction and model training, a malware classifier is generated. When a sample is input into the classifier, the classification result and a list of features with different weights are obtained. We remove those features that do not exist in the sample and sort the left features according to their weights. Then we select the top $n$ features to generate the behavior description.
Here, $n$ is a \revised{hyperparameter}. 
It is important to select a proper number for $n$. Although choosing more features as key features may help to identify more malicious behaviors, too many features will reduce the interpretability of classification~\citep{ribeiro2016should}. 
The number for $n$ is a heuristic value depending on concrete scenarios. According to the experiments, the default value is configured as 6.

{Our customized model utilizes a fully connected network in the attention layer to capture the correlation between features, rather than a multi-layer fully connected network. A multi-layer fully connected network may capture much more complex relationships between features, but it is also difficult to understand and interpret {since it involves too many mathematical operations, 
{making} it impossible for humans to follow the exact mapping from input feature to output}. 
That is one reason why we do not use CNN or RNN models. 
{In general,} the deep learning models still cannot be interpreted accurately. How to interpret deep neural networks is an open challenge so far, which 
{also belongs }to our future work.}

\subsection{Malware Description Generation}\label{ssec:description}

\begin{figure}\centering
\includegraphics[width=0.95\textwidth]{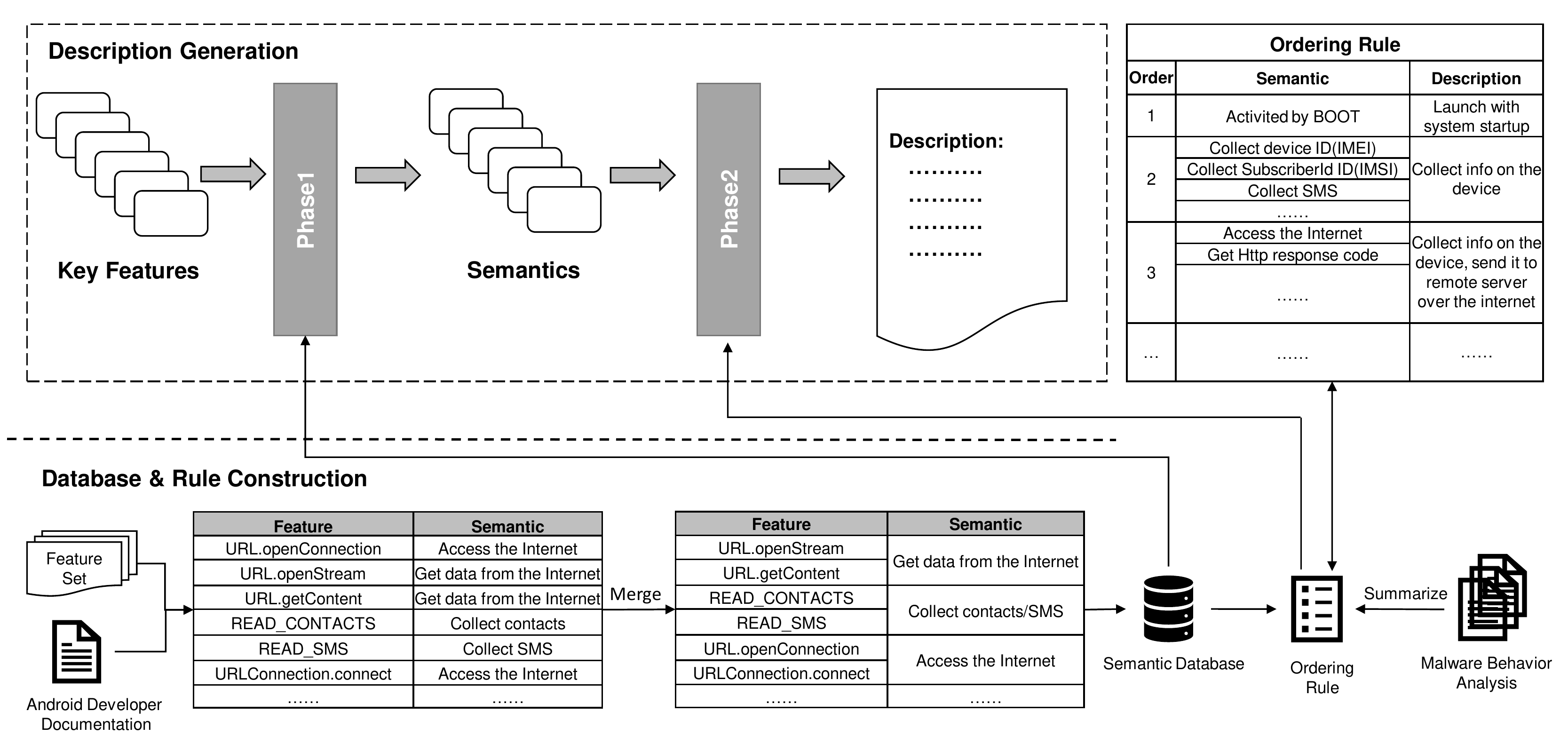}
\caption{\revised{Malware description generation}}
\label{fig:interpreter}
\end{figure}

In order to generate malicious behavior description for Android malware, \revised{we first match the malware key features to their corresponding semantics.} 
We select a 158-dimensional feature vector as input to train the classifiers. We search Android developer documentation \citep{ref_url3} for the semantics of each feature according to 
{its name.}
The Android developer documentation has a detailed functional description for each API and permission. We download the detailed functional description of each feature.
However, the functional descriptions 
include too many details and are difficult to understand 
{comprehensively}. \revised{We simplify and generalize them into simple semantics by intercepting and generalizing the key predicates, objects, and complements}. \revised{For example, the functional description of READ\_CONTACTS is ``Allows an application to read the user's contacts data''. We generalize it as ``Collect contacts''. 
Similarly, permission.READ\_CONTACTS is generalized as ``Collect contacts''.
After that, we use the feature and the corresponding semantic to build a semantic database (shown in Fig.~\ref{fig:interpreter}).}
According to our observation, {some features share the same semantics}. For instance, URL.openconnection and URLConnection.connect share the same semantics of \qq{Access the Internet}. 
Besides, 
\revised{some features exhibit a similar functionality and can be combined into one semantic feature. For example, permission.READ\_CONTACTS and permission.READ\_SMS are both about information collection and can be combined into \qq{Collect contact/SMS}.}
Therefore, we denote two rules as follows:

\begin{itemize}
\item \textit{\textbf{Rule 1:} If features {belong to a same functionality}, they are assigned the same semantics. } 
\item  \textit{\textbf{Rule 2:} If features {exhibit a similar functionality}, they are assigned the similar semantics and the two similar semantics are combined into one. }
\end{itemize}

In this way, we match features with semantics based on their functional descriptions so as to obtain simple and useful semantics for features. After that, we convert the semantics into malware descriptions to make it easier for users to understand. In order to generate reasonable descriptions for the Android malware, 
\revised{we summarize ten basic malicious behaviors from a large number of malware, and establish the mapping relation between the malicious behaviors and their corresponding semantics. We also define some ordering rules to arrange the semantics (shown in Fig.~\ref{fig:interpreter}, ordering rule) according to the malware behavior analysis by manual (shown in Fig.~\ref{fig:interpreter}, malware behavior analysis).}
For example, if ``activated by BOOT'' exists, it should be ranked first; If 
{``access the internet''}and ``collect IMEI'' exist at the same time, ``collect IMEI'' should be in front of ``Access the internet''. Therefore, when ``access to the Internet", ``collect IMEI'' and ``Activate by BOOT'' exist simultaneously, the order should be ``activated by BOOT'', ``collect IMEI'' and ``access the Internet''. Then they are converted into ``Launch with system startup, collect info on the device, and send it to remote server over the Internet'' \revised{through the mapping relation between semantics and malware behaviors}. 

Specifically, we first get a set of key features $U$, where $k_i \in U$ is the $ i$-th key feature. Then we converted $k_i$ into $s_i$ one by one \revised{in the phase1 shown in Fig.~\ref{fig:interpreter}}, where $s_i$ is the $ i$-th semantics. According to Rule 1, if key features belong to a same functionality, they are assigned to the same semantics. Therefore, those semantics that exist in $S$ are not added to $S$ again. According to Rule 2, if the key features exhibit a similar functionality, their similar semantics are combined into one. Therefore, when the semantics $s_i$ similar to semantic $s$ in $S$ appears, we combine it with $s$ and then update $s$ in $S$.  After that, we convert the semantics into descriptions one by one \revised{in the phase2 shown in Fig.~\ref{fig:interpreter}}. \revised{Fig.~\ref{fig:interpreter} shows how the interpreter generates the malware description step by step and how the semantic database and ordering rule are established.}
\revised{The implementation details of the semantic database and ordering rules are provided on our website:~{\url{https://sites.google.com/view/xmal/}}.}

{\section{Evaluation}}

In this paper, we aim to utilize the proposed method \mytool to explain why an app is classified as malware. 
However, before interpreting the classification results, we should ensure that the detection accuracy is high enough since the malware detection accuracy is as important as the interpretability results, otherwise,
the interpretation is meaningless. Therefore, in this section, we perform experiments to evaluate the malware detection accuracy and interpretability of the proposed method. 
\revised{Additionally, we also conduct an in-depth comparison study between \mytool and the state-of-the-art techniques.
We aim to answer the following research questions in our evaluation.}

\vspace{1mm}
\subsection{RQ1: What is the detection accuracy of \mytool in the malware classification?}

\revised{In this experiment, we first investigate the Android malware detection performance of \mytool. We adapt the best {hyperparameters} of \mytool for the best detection performance, and then conduct experiments to evaluate \mytool and compare it with the state-of-the-art techniques. Finally, we 
investigate whether \mytool can {further} be extended to the unsupervised Android applications in the wild.}

\subsubsection{Dataset.}\label{sec:dataset}
To conduct the experiment, we first collect a large amount of Android malware from two sources:
{10,010} samples from the \textit{National Internet Emergency Center}~\citep{cert}  and {5,560}
samples from Drebin~\citep{arp2014drebin}. Most of the samples from the National Internet Emergency Center are the recent malicious samples rather than from old datasets such as Gemome~\citep{zhou2012dissecting} in 2011. These malware samples include {a variety of}
threats for Android, such as data leakage, phishing, trojans, spyware, and root exploits. 
\revised{Apart from these malicious apps, we also fetched the top apps overall per category from Google Play Store {and HUAWEI app store} on July 2019 and collect {20,193}
apps in total.}
We removed the ones that are classified as Android malware candidates by VirusTotal service~\citep{virustotal}. Finally, we obtain
{20,120}
benign samples in total and {15,570}
malicious samples, which are available on our website~{\url{https://sites.google.com/view/xmal/}.}

\subsubsection{Setup.} \label{subsec:procedure1}
To select the best hyperparameters for \mytool, we first randomly split the {15,570} Android malware samples and 20,120 benign apps into a training set (\revised{70\%, i.e., {24,983}
samples in total}) and a test set (\revised{30\%, i.e., {10,707}
samples in total}). \revised{Note that these two sets have no overlap in our experiments.} After that, we extract 158-dimensional feature vectors including 97 API calls and 61 permissions from the training set to train \mytool, and  utilize test set to evaluate the detection accuracy of \mytool.
\revised{Then we test different hyperparameters to \mytool and finally determine the hyperparameters that can achieve the best detection performance.} 

In order to further demonstrate the superiority of \mytool, we compare it with the state-of-the-art techniques and use \textit{recall, precision, accuracy,} and \textit{F-measure} to evaluate the detection performance. In this experiment, we compare \mytool with Drebin \citep{arp2014drebin} and LIME \citep{ribeiro2016should}. 
\revised{The reasons we select these two methods are as follows: 
1) Drebin is a model-specific interpretable method like \mytool, and achieve a high Android malware detection accuracy of 93.90\%. 
2) LIME proposes an effective model-agnostic method to interpret individual model prediction and obtain a high classification accuracy. It is one of the most valuable methods in model interpretability and has been cited and compared by many related research studies~\citep{guo2018lemna,lundberg2017unified,samek2017explainable,montavon2018methods,miller2019explanation}.
Comparing \mytool with LIME is very significant for evaluating the interpretability and detection accuracy of \mytool.} 

\revised{
{In conducting a comparison experiment}, we first re-implement Drebin based on its published research paper. Since LIME is open-source, we are able to reuse it directly for our experiments.} 
Specifically, for Drebin, we extract 422-dimensional features including API, permission, intent, activity, service, and hardware components from the dataset, and utilize them to train and test the model in Drebin. 
For LIME, since it is a model-agnostic method, we apply it to the MLP model. We extract the same features as \mytool from the data set mentioned in the accuracy experiment to train and test the MLP model. 

\revised{Moreover, to evaluate the detection accuracy of \mytool in different malware families, we also select the top 16 malware families with the largest number of samples (cf. Table \ref{tb4}) according to the malware family tags provided by Drebin~\citep{arp2014drebin}. Since some malware families have too few samples to validate the interpretable results in the next experiments, we randomly select 10 samples for each malware family (i.e., 160 samples in total) {from the test set} 
for further investigation. 
\revised{In addition, in order to validate the detection accuracy of \mytool in the samples from National Internet Emergency Center, we also randomly select 10 malicious samples {from the test set}.}
Finally, 170 malicious samples are selected to test and compare, we also randomly select 170 benign apps from the test set accordingly to validate \mytool.}

\begin{figure}
  \centering 
  \subfigure[\revised{Optimizer \& Activation}]{ 
    \label{fig:exp1:a}
    \includegraphics[width=0.48\textwidth]{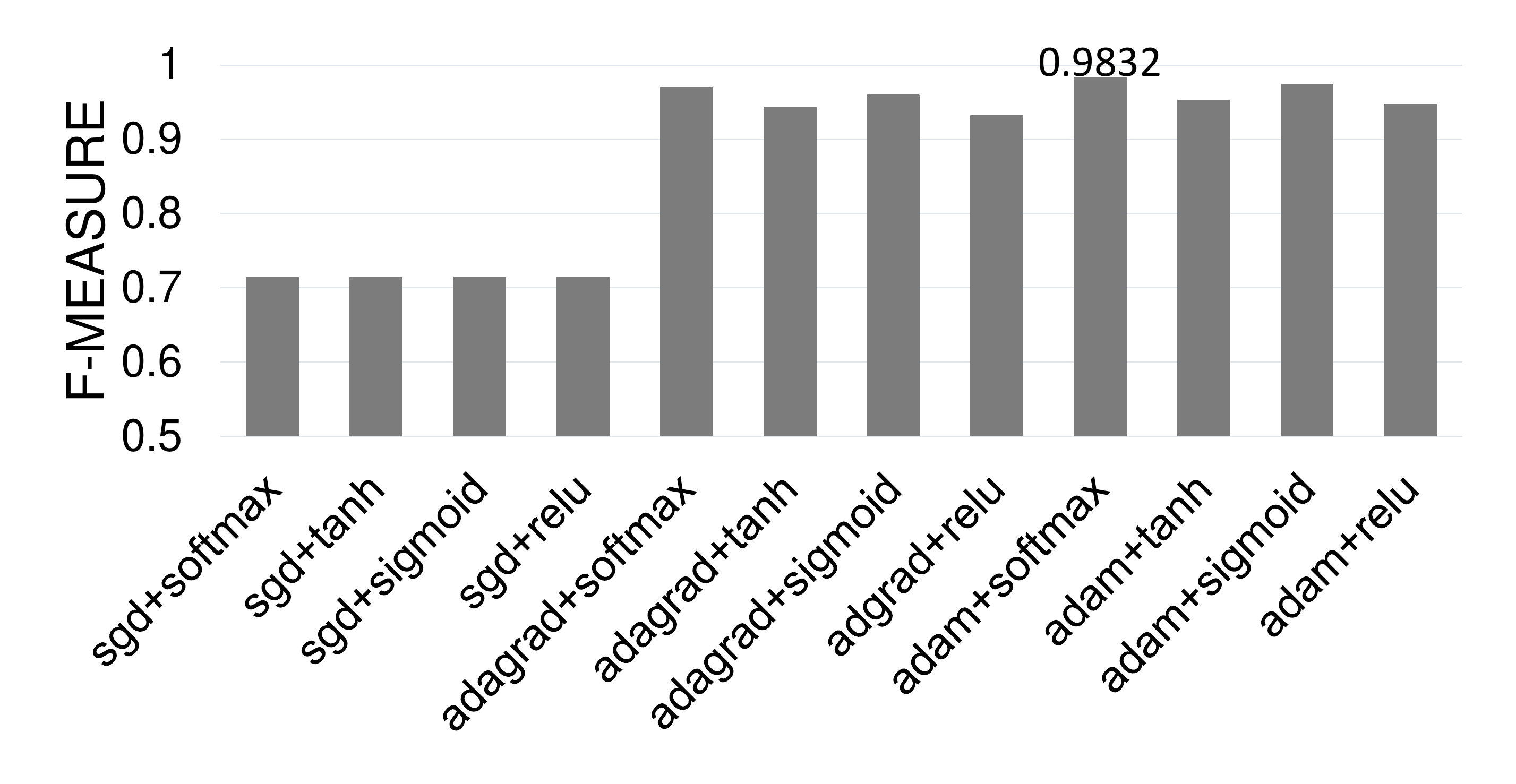} 
  } 
  \subfigure[\revised{Epochs \& Batch\_size}]{ 
    \label{fig:exp1:b}
    \includegraphics[width=0.48\textwidth]{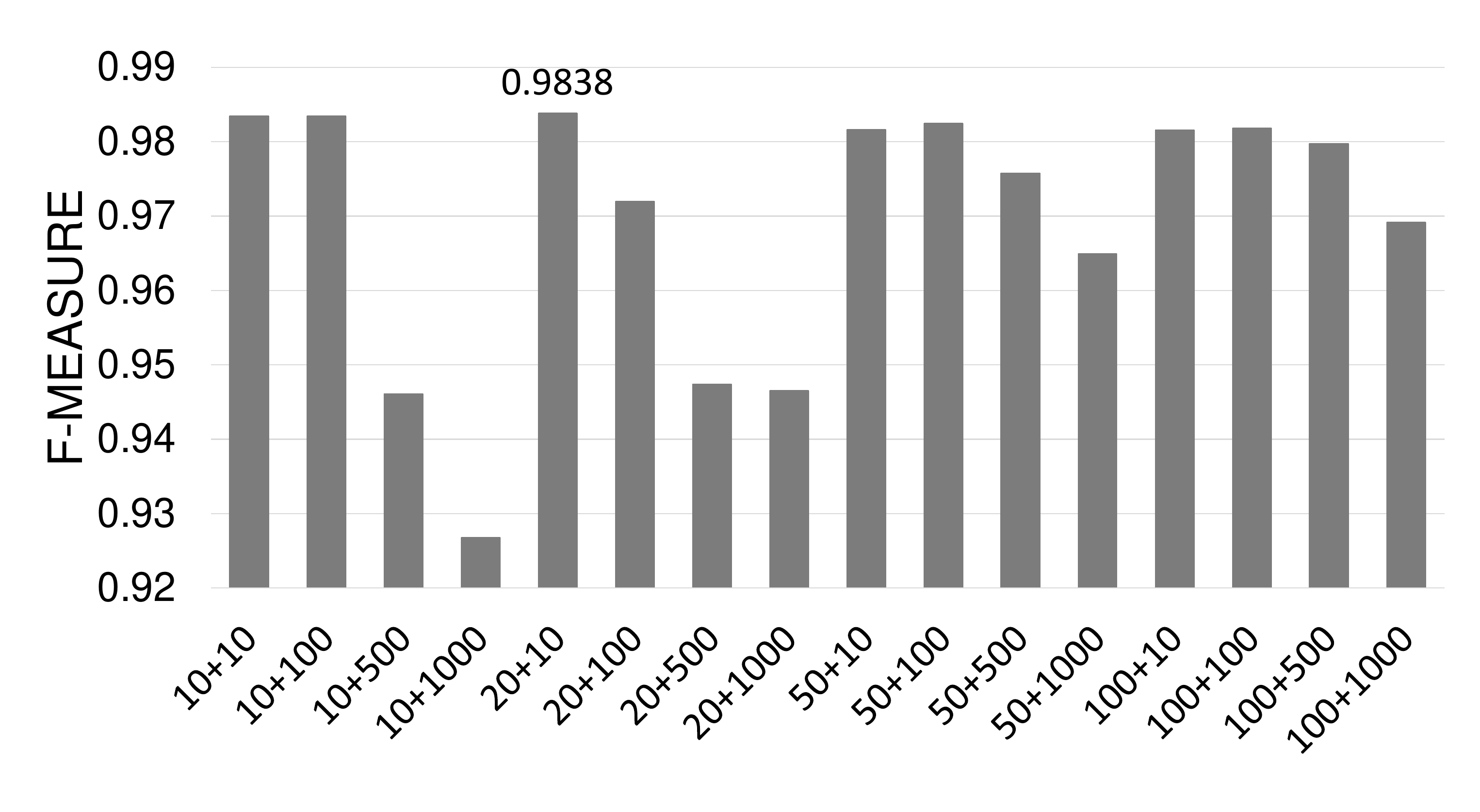} 
  } 
  \caption{\revised{Detection precision and recall under the different hyperparameters}} 
  \label{fig:exp1} 
\end{figure}

\vspace{1mm}
\noindent \textbf{Parameter tuning for best classification performance.}
\revised{To achieve a better detection performance, we first search for the best hyperparameters (i.e., learning rate, optimizer, activation function, epochs, and batch\_size) of \mytool. Specifically, we set the learning rate to a set of values including 0.0001, 0.001, 0.01, and 0.1, which shows a little difference in detection performance. Therefore, we select 0.001 as the learning rate in our experiments. Fig.~\ref{fig:exp1:a} demonstrates the detection results of applying different optimizer and activation function. As a result, \qq{adam + softmax} achieves the best performance overall. We further investigate the impact of epochs and batch\_size. As shown in Fig.~\ref{fig:exp1:b}, the configuration of 10 epochs and 20 batch\_size achieves the best result.}

\subsubsection{Results.}\label{subsec:accuracy}

\begin{table}[t]\small
\centering
\caption{{Detection results of three models}}
\begin{tabular}{lccc}
\toprule
\textbf{Models}  & \textbf{Drebin} & \textbf{MLP in LIME} & \textbf{\mytool} \\ \midrule
\textbf{Recall}    & 94.90\%         & 97.13\%              & 98.28\%                         \\
\textbf{Precision}    & 95.94\%         & 96.38\%              & 98.48\%                         \\
\textbf{Accuracy}    & 95.24\%         & 96.50\%              & 98.35\%                         \\
\textbf{F-measure}    & 95.42\%         & 96.75\%              & 98.37\%                         \\
\bottomrule
\end{tabular}
\label{tb22}
\end{table}

\revised{After determining the best hyperparameters of \mytool, we use the test set ({10,707} samples in total) to test \mytool, Drebin, and LIME, and compare their detection performance. The experiment results are shown in Table \ref{tb22}. The result shows that the accuracy of the re-implemented model of Drebin is 95.24\%, while the original accuracy in the paper is 93.90\%, indicating the model we implemented is comparable to the original model. Note that, the accuracy of the re-implemented model is better than the original model because we perform the feature selection by using manual statistical pruning used in~\citep{chen2018automated}.}
The accuracy of the three models are all above 95\%, while \mytool achieves 98.35\% detection accuracy, outperforms the other two methods. Moreover, we evaluate \mytool on the 170 benign test sample with a TNR (true negative rate) of 98.82\%, which means that only 2 benign applications are misclassified as malware. We also test \mytool on the 170 malware samples. The TPR (true positive rate) of the 10 malware samples from National Internet Emergency Center is 100\%, and the TPR of each family is shown in Table 3. We can see that most malware families have a TPR of 100\%, while DroidKungFu and SMSreg have TPRs of 80\% and 90\%, respectively, which means only two DroidKungFu malware and one SMSreg malware are misclassified as benign. In summary, \mytool achieves high detection accuracy in malware detection. \revised{In order to validate \mytool on unsupervised cases, we also randomly collect {1,000} Android apps from several Android application markets (e.g., Google Play Store, APKpure, coolapk, appchina, and mi.com) and employ \mytool on these samples. We find that {five of them} are classified as malware. After manual analysis, we confirm that three apps\footnote{SHA1 values: 7BA69225D0B9B06DCADACA693DF58DE03228CDBE, AEDCB0B03C9193AC1F4B9CCFB31DDBA0FB7D9510, and AA0A1B157EA57E753C793F68141155A5A72F0620} privately obtain the users' contacts and send them to a malicious remote server. Now they have been removed from the app stores. {The other two apps\footnote{SHA1 values: 4246c467eb833805a0e7c09df0e8d72cf182bdfa and 8075a71fd8165fd1e33652fd7cd55f06b09a1697} trick users into downloading apps and collect users' information. They still can be found on the website~{\url{http://www.appchina.com/soft}} and we will report accordingly.}}

\begin{table}[t]\small
\centering
\caption{Detection accuracy of the top 16 malware families}
\begin{tabular}{lc|lc|lc|lc}
\toprule
\textbf{Families}  & \textbf{TPR} & \textbf{Families}  & \textbf{TPR} & \textbf{Families}  & \textbf{TPR} & \textbf{Families}  & \textbf{TPR} \\ \midrule
Adrd & 100\% & DroidKungFu & 80\% & BaseBridge & 100\% & Geinimi & 100\% \\ \midrule
DroidDream & 100\% & GinMaster & 100\% & SendPay & 100\% & Iconosys & 100\% \\ \midrule
FakeDoc & 100\% & Gappusin & 100\% & Plankton & 100\% & Kmin & 100\% \\ \midrule
FakeInstaller & 100\% & MobileTx & 100\% & SMSreg & 90\% & Opfake & 100\% \\ \bottomrule
\end{tabular}
\label{tb4}
\end{table}


\subsection{\revised{RQ2: How well does the malware description generated by \mytool match the actual malicious behaviors of the malware?}}

\revised{We aim to interpret why an app is classified as malware in this section.
To answer it, we conduct experiments to investigate whether the malware descriptions can match the actual malicious behavior of the malware.}

\subsubsection{Dataset.}


\revised{We perform interpretability experiments on all malicious samples ({15,570} in total collected in Section \ref{sec:dataset}) and generate the corresponding malware descriptions for each of them. In order to evaluate whether the malware descriptions generated by \mytool match the actual malicious behaviors of the malware, we use the 170 malicious samples mentioned in Section~\ref{subsec:procedure1} to 
{establish} the ground truth.}
Among them, since the 10 malware samples from National Internet Emergency Center have been analyzed before and the corresponding expert analysis has been validated by 
{the} expert team, 
we directly 
{employ} them as the ground truth of malware description. 
\revised{Note that, all members of the expert team are from National Internet Emergency Center and have engaged in malware analysis on the platform of Android and 
{Windows} for more than 3 years. They perform malware analysis and computer forensics on a daily basis, and are good at analyzing the malicious code/behaviors and identifying malware manually. For the other 160 samples from the top 16 malware families, we collect the corresponding expert analysis reports of each family from Symantec~\citep{stnabtec} and Microsoft~\citep{Microsoft}. Meanwhile, we collaborate with {the} experienced expert team from National Internet Emergency Center, {where} they manually analyze these malware samples and provide the corresponding analysis reports. 
After that, we cross-validate the analysis reports from the two different resources and obtain the final ground truth of malware descriptions.
{Consequently,} we can evaluate the interpretability results by comparing with the ground truth.} 

\revised{In addition, to further evaluate whether \mytool can explain why the benign application is misclassified
as malicious, we also select {the} 170 benign apps mentioned in Section~\ref{subsec:procedure1} to conduct the experiments.}


\subsubsection{Setup.}\label{subsec:procedure2}
\revised{We first use \mytool to generate the malicious behavior descriptions of the {15,570} malicious samples and evaluate 
{if} the generated descriptions (170 malware samples) match with the ground truth. In addition, to further evaluate whether XMal can explain why the benign application is misclassified as malicious, we 
employ \mytool on the 170 benign samples to conduct 
{more} experiments.}
\revised{
However, the evaluation may be biased by our subjective opinions. 
In order to mitigate this problem, we randomly select one sample from each malware family and conduct an online survey to investigate the quality of the malware description generated for these samples. Moreover, we also conduct a quantitative analysis to validate the effectiveness of \mytool. 
}

\vspace{1mm}
\noindent \textbf{Evaluation Metrics.} 
\revised{The ground truth and generated malware descriptions by \mytool are unstructured text, which cannot be compared quantitatively. Inspired by Grounded Theory ~\citep{corbin2014basics, stol2016grounded}, we extract  \qq{{concepts}} from the ground truth and 
{generate} malware descriptions, and compute how many  \qq{{concepts}} in the ground truth can be detected by \mytool and how many \qq{{concepts}} in the generated descriptions do not exist in the ground truth. Here, \qq{concept} refers to a meta-behavior. For instance, \qq{Activate when the mobile device is booted up} consists of two \qq{{concepts}}, \qq{Activate} and \qq{the mobile device is booted up}. Here, we let \textit{total\_concepts} be the total number of  \qq{{concepts}} in the ground truth, \textit{detect\_concepts} be the number of \qq{{concepts}} in the ground truth that are detected by \mytool, and \textit{surplus\_concepts} be the number of \qq{{concepts}} in the generated descriptions that do not exist in the ground truth. In order to quantitatively measure the interpretability results, we define the evaluation metric \qq{interpretability result} (a.k.a. \textit{ir}) as follows, and use \textit{ir} to evaluate the generated descriptions of all malware by \mytool.
\begin{equation}
    precision  = \frac {detect\_concepts}{detect\_concepts + surplus\_concept}
\end{equation}
\begin{equation}
    recall  = \frac {detect\_concepts}{total\_concepts}
\end{equation}
\begin{equation}
    ir = \frac {2 \times precision \times recall}{precision + recall}
\end{equation}}

\revised{As {the number of} \textit{detect\_concepts} increases, \textit{ir} becomes larger. When {the number of} \textit{surplus\_concepts} increases, \textit{ir} becomes smaller. Therefore, the closer \textit{ir} is to 1, the better the interpretability result. We take the \textit{Adrd} in Table~\ref{tb5} as an example to illustrate the calculation process of \textit{ir}. We extract concepts from the ground truth and the generated description and 
{list} them in 
Table ~\ref{tb20}. Specifically, {\qq{activate}} 
{vs. }\qq{launch} are the same concept. Similarly, \qq{the mobile device is booted up} 
{vs. }\qq{system startup}, {\qq{access the Internet} 
{vs.} \qq{over the Internet}}, \qq{stead some info} 
{vs.} \qq{collect info on the device}, {and} \qq{send to remote server} 
{vs.} \qq{send it to remote server} are also the same concept. Consequently, we can know that \textit{detect\_concepts} is {5}, \textit{surplus\_concepts} is 0, and \textit{total\_concepts} is {6}. So \textit{ir} is {0.91}.} 

\begin{table}\small
\caption{\revised{The \qq{Concepts} of {Adrd}}}
\begin{tabular}{ll}
\toprule 
 & \textbf{Concepts} \\ \hline
\textbf{Ground Truth} &
  \begin{tabular}[c]{@{}l@{}}{1. activate}  \  2. the mobile device is booted up \ {3. access the Internet} \\{4. download components} 5. stead some info 6. send to remote server \end{tabular} \\ \hline
\textbf{Generated Description} &
  \begin{tabular}[c]{@{}l@{}}1. launch  \ 2. system startup  \ 3. collect info on the device  \\ 4. send it to remote server  \ 5. over the internet \end{tabular} \\ \bottomrule 
\end{tabular}
\label{tb20}
\end{table}

\vspace{1mm}
\noindent \textbf{Parameter tuning for best interpretability.}
\revised{Before conducting the experiments, we perform hyperparameter tuning to select a proper value for \textit{n}. We first obtain 50 malware samples and {the} corresponding expert analysis reports from National Internet Emergency Center. Then, we evaluate \mytool and calculate \textit{ir} of all samples under different values of \textit{n} {(i.e., ranging from 1 to 10). The results are shown in Table ~\ref{tb90}.} We find that when \textit{n} is set to 6, the average of \textit{ir} for all samples is closest to 1, which is 0.92. Therefore, in the following experiments, \textit{n} is set to 6. Note that 6 is not 
the best number of features in all scenarios. Hyperparameter tunning is necessary for different scenarios.}

\begin{table*}[t]
\centering
\caption{\revised{The \textit{ir} computed by different values of \textit{n}}}
\begin{tabular}{c|c|c|c|c|c|c|c|c|c|c}
\hline
\textbf{n}  & 1   & 2    & 3    & 4    & 5    & 6    & 7    & 8    & 9    & 10   \\ \hline
\textbf{ir} & 0.45 & 0.67 & 0.68 & 0.75 & 0.87 & 0.92 & 0.72 & 0.69 & 0.62 & 0.58 \\ \hline
\end{tabular}
\label{tb90}
\end{table*}

\subsubsection{Results.}
We select one sample from each malware family and two samples (named ``blackgame'' and ``xunbaikew1'') from National Internet Emergency Center to demonstrate the interpretability of \mytool. The interpretability results are as shown in Table~\ref{tb5}. To illustrate how the experimental results explain why an app is classified as malware, we take Adrd and Opfake families as examples.

\begin{table*}
\centering 
\footnotesize
\caption{Part of the Interpretability Results of \mytool. 
The full list can be found on our website~{\url{https://sites.google.com/view/xmal/}}}
\scalebox{0.75}{
\begin{tabular}{cllll}
\toprule
\multicolumn{1}{c}{\textbf{{}}} & \multicolumn{1}{l}{\textbf{Key Features}} & \multicolumn{1}{l}{\textbf{Semantics Matching}} & \multicolumn{1}{l}{\textbf{Description Generated by \mytool}} & \multicolumn{1}{l}{\textbf{Expert Analysis (Ground Truth)}} \\ \midrule
\multirow{6}{*}{\rotatebox{90}{Adrd}} & \multirow{6}{*}{\begin{tabular}[c]{@{}l@{}}URL.openConnection\\ READ\_PHONE\_STATE\\ RECEIVE\_BOOT\_COMPLETED\\ {LocationManager.requestLocationUpdates}\\ {HttpURLConnection.getResponseCode}\\ TelephonyManager.getSubscriberId\end{tabular}} & \multirow{6}{*}{\begin{tabular}[c]{@{}l@{}}1. Access the Internet\\ 2. Collect IMEI/IMSI/location\\ 3. Activated by BOOT\end{tabular}} & \multirow{6}{*}{\begin{tabular}[c]{@{}l@{}}Launch with system startup, \\collect info on the device,\\ and send it to remote server \\ over the internet\end{tabular}} & \multirow{6}{*}{\begin{tabular}[c]{@{}l@{}}1. {Activate} when the mobile\\device is booted up.\\{2. Access the Internet and}\\{download components} \\3. Steal some info and send to \\remote server.\end{tabular}} \\
\multicolumn{1}{c}{} &  &  &  &  \\
\multicolumn{1}{c}{} &  &  &  &  \\
\multicolumn{1}{c}{} &  &  &  &  \\
\multicolumn{1}{c}{} &  &  &  &  \\
\multicolumn{1}{c}{} &  &  &  &  \\ \midrule
\multirow{6}{*}{\rotatebox{90}{BaseBridge}} & \multirow{6}{*}{\begin{tabular}[c]{@{}l@{}}SEND\_SMS\\URL.openConnection\\ READ\_PHONE\_STATE\\ RECEIVE\_SMS\\URLConnection.connect\\  {RECEIVE\_BOOT\_COMPLETED}\end{tabular}} & \multirow{6}{*}{\begin{tabular}[c]{@{}l@{}}1. Send SMS messages\\ 2. Access the Internet\\ 3. {Collect IMEI/SMS}\\  {4. Activated by BOOT}\end{tabular}} & \multirow{6}{*}{\begin{tabular}[c]{@{}l@{}} {Launch with system startup}, \\send SMS to premium-rate \\numbers, collect info on the \\device, and send it to remote \\server over the internet\end{tabular}} & \multirow{6}{*}{\begin{tabular}[c]{@{}l@{}}1. Send and receive SMS\\ 2. {Info is sent to remote server}:\\ a) Subscriber ID \\ b) Device manufacturer/model\\ c) Android OS version\\3. {Activate when the mobile starts}\end{tabular}} \\
 &  &  &  &  \\
 &  &  &  &  \\
 &  &  &  &  \\
 &  &  &  &  \\
 &  &  &  &  \\ \midrule
\multicolumn{1}{c}{\multirow{6}{*}{\rotatebox{90}{DroidKungFu}}} & \multirow{6}{*}{\begin{tabular}[c]{@{}l@{}}URL.openConnection\\ READ\_EXTERNAL\_STORAGE\\ READ\_PHONE\_STATE\\ URLConnection.getURL\\ URLConnection.connect\\  {RECEIVE\_BOOT\_COMPLETED}\end{tabular}} & \multirow{6}{*}{\begin{tabular}[c]{@{}l@{}}1. Access the Internet\\ 2. Write to external storage\\ 3. Collect IMEI\\ 4.  {Activated by BOOT}\end{tabular}} & \multirow{6}{*}{\begin{tabular}[c]{@{}l@{}} {Launch with system startup}, \\download malware to SD card, \\collect info on the device, \\and send it to remote server\\ over the internet\end{tabular}} & \multirow{6}{*}{\begin{tabular}[c]{@{}l@{}}1. Steal sensitive info: \\  IMEI number, device version, \\ operating system version, etc.\\ 2. Download files from remote \\computer or the internet.\\\end{tabular}} \\
\multicolumn{1}{c}{} &  &  &  &  \\
\multicolumn{1}{c}{} &  &  &  &  \\
\multicolumn{1}{c}{} &  &  &  &  \\
\multicolumn{1}{c}{} &  &  &  &  \\
\multicolumn{1}{c}{} &  &  &  &  \\ \midrule
\multirow{6}{*}{\rotatebox{90}{FakeInstaller}} & \multirow{6}{*}{\begin{tabular}[c]{@{}l@{}}SEND\_SMS\\ READ\_PHONE\_STATE\\ RECEIVE\_SMS\\ READ\_SMS\\  {TelephonyManager.getNetworkOperator}\\  {WAKE\_LOCK}\end{tabular}} & \multirow{6}{*}{\begin{tabular}[c]{@{}l@{}}1. Send SMS messages\\  {2. Collect IMEI/SMS}\\  {3. Unlock phone}\end{tabular}} & \multirow{6}{*}{\begin{tabular}[c]{@{}l@{}}Send SMS to premium-rate\\ numbers,  {collect info on the}\\ {device, keep running in the} \\  {background}\end{tabular}} & \multirow{6}{*}{\begin{tabular}[c]{@{}l@{}}1. Send the premium SMS\\2. Receive commands from \\{a remote server} \end{tabular}} \\
 &  &  &  &  \\
 &  &  &  &  \\
 &  &  &  &  \\
 &  &  &  &  \\
 &  &  &  &  \\ \midrule
\multirow{5}{*}{\rotatebox{90}{Gappusin}} & \multirow{5}{*}{\begin{tabular}[c]{@{}l@{}}URL.openConnection\\ READ\_PHONE\_STATE\\ RECEIVE\_BOOT\_COMPLETED\\ NotificationManager.notify\end{tabular}} & \multirow{5}{*}{\begin{tabular}[c]{@{}l@{}}1. Access the Internet\\ 2. Collect IMEI\\ 3. Activated by BOOT\\ 4. Notify the info\end{tabular}} & \multirow{5}{*}{\begin{tabular}[c]{@{}l@{}}Launch with system startup, \\collect info on the device, \\and send it to remote server\\ over the Internet, send a\\ notification as system\end{tabular}} & \multirow{5}{*}{\begin{tabular}[c]{@{}l@{}}1. Post device info such as \\IMEI, IMSI, and OS version.\\ 2. Download apps/disguises \\ as system updates.\end{tabular}} \\
 &  &  &  &  \\
 &  &  &  &  \\
  &  &  &  &  \\
 &  &  &  &  \\ \midrule
 \multirow{4}{*}{\rotatebox{90}{Opfake}} & \multirow{4}{*}{\begin{tabular}[c]{@{}l@{}}SEND\_SMS\\ URL.openConnection\\ READ\_PHONE\_STATE\\ {TelephonyManager.getNetworkOperator}\end{tabular}} & \multirow{4}{*}{\begin{tabular}[c]{@{}l@{}}1. Send SMS messages\\ 2. Access the Internet\\ 3. Collect {IMEI}\end{tabular}} & \multirow{4}{*}{\begin{tabular}[c]{@{}l@{}}Send SMS to premium-rate \\numbers, collect info on the \\device, and send it to remote \\server over the Internet\end{tabular}} & \multirow{4}{*}{\begin{tabular}[c]{@{}l@{}}1. {Send SMS to premium-rate num.}\\ 2. Access info about network.\\ 3. {Check the phone's current state.}\end{tabular}} \\
 &  &  &  &  \\
 &  &  &  &  \\
 &  &  &  &  \\ \midrule

\multirow{6}{*}{\rotatebox{90}{blackgame}} & \multirow{6}{*}{\begin{tabular}[c]{@{}l@{}}URL.openConnection\\ SEND\_SMS\\RECEIVE\_SMS\\WRITE\_SMS\\TelephonyManager.getDeviceId\\TelephonyManager.getSubscriberId\end{tabular}} & \multirow{6}{*}{\begin{tabular}[c]{@{}l@{}}1. Access the Internet\\ 2. Send SMS messages\\ 3. Collect SMS/IMEI/IMSI\end{tabular}} & \multirow{6}{*}{\begin{tabular}[c]{@{}l@{}}Send SMS to premium-rate\\ numbers, collect info on the\\ device, and send it to remote\\ server over the internet\end{tabular}} & \multirow{6}{*}{\begin{tabular}[c]{@{}l@{}}1. Send SMS to premium-rate num.\\ 2. obtain phone num and device info \\and upload it to the remote server.\end{tabular}} \\
\multicolumn{1}{c}{} &  &  &  &  \\
\multicolumn{1}{c}{} &  &  &  &  \\
\multicolumn{1}{c}{} &  &  &  &  \\
\multicolumn{1}{c}{} &  &  &  &  \\
\multicolumn{1}{c}{} &  &  &  &  \\ \midrule

\multirow{6}{*}{\rotatebox{90}{xunbaikew1}} & \multirow{6}{*}{\begin{tabular}[c]{@{}l@{}}SEND\_SMS\\ ContentResolver.query\\READ\_CONTACTS\end{tabular}} & \multirow{6}{*}{\begin{tabular}[c]{@{}l@{}}1. Send SMS messages\\ 2. Collect contact info\end{tabular}} & \multirow{6}{*}{\begin{tabular}[c]{@{}l@{}}Collect contact info on the \\device, and  {send SMS to}\\ {premium-rate num}\end{tabular}} & \multirow{6}{*}{\begin{tabular}[c]{@{}l@{}} Collect contact info, and then send \\SMS with the app download link \\to all contacts.\end{tabular}} \\
\multicolumn{1}{c}{} &  &  &  &  \\
\multicolumn{1}{c}{} &  &  &  &  \\
\multicolumn{1}{c}{} &  &  &  &  \\
\multicolumn{1}{c}{} &  &  &  &  \\
\multicolumn{1}{c}{} &  &  &  &  \\ \midrule

\end{tabular}
}
\label{tb5}
\end{table*}

\begin{table}\footnotesize
\centering
\caption{Two misclassified benign apps}
\scalebox{0.9}{
\begin{tabular}{lll}
\toprule
\textbf{Sample} & \textbf{Key Features} & \textbf{Semantics Matching} \\ \midrule
\multirow{4}{*}{HiViewTunnel} & \multirow{4}{*}{\begin{tabular}[c]{@{}l@{}}permission.INTERNET\\ {WRITE\_EXTERNAL\_STORAGE} \\ URL.openConnection\\ {TelephonyManager.getDeviceId}\end{tabular}} & \multirow{4}{*}{\begin{tabular}[c]{@{}l@{}}1. Access the Internet\\ 2. Write to external storage\\ 3. Collect DeviceId\end{tabular}} \\
 &  &  \\
 &  &  \\
 &  &  \\ \midrule
\multirow{5}{*}{HwSpaceService} & \multirow{5}{*}{\begin{tabular}[c]{@{}l@{}}permission.INTERNET\\ URL.openConnection\\ {WAKE\_LOCK}\\ ContentResolver.query\\ {READ\_PHONE\_STATE}\end{tabular}} & \multirow{5}{*}{\begin{tabular}[c]{@{}l@{}}1. Access the Internet \\ 2. Unlock phone\\ 3. Collect SMS/IMEI\end{tabular}} \\
 &  &  \\
 &  &  \\
 &  &  \\
 &  &  \\ \bottomrule
\end{tabular}
}
\label{tb6}
\end{table}
Android.Adrd is {a Trojan horse in Adrd malware family} that steals information from Android devices. As shown in Table \ref{tb5}, \mytool outputs 6 key features (i.e., URL.openConnection, READ\_PHONE\_STATE, RECEIVE\_BOOT\_COMPLETED, requestLocationUpdates, getResponseCode, and getSubscriberId) for a sample of Adrd, and generates the corresponding semantics (i.e., ``Access the Internet'', ``Collect IMEI/IMSI/location'', and ``Activate by BOOT'') and malicious behavior description (i.e, \textit{``Launch with system startup, collect info on the device, and send it to remote server over the internet''}). The malicious behavior description generated by \mytool can clearly explain the reason why the sample of Adrd is classified as malware. In addition, the expert analysis of Adrd in Table \ref{tb5} shows that it has the behavior of re-executing itself when the mobile device is booted up, stealing information and sending to a remote server. This is consistent with the malicious behavior description generated by \mytool, which demonstrates the effectiveness of \mytool. Additionally, we cross-validate through three co-authors to determine whether the semantics of the generated description by \mytool is consistent with the ground truth (i.e., Expert Analysis). We accept the result only if all of us agree on it.

Opfake family sends SMS messages to premium-rate numbers on the Android platform. As can be seen from Table~\ref{tb5}, \mytool outputs four key feature (i.e., {SEND\_SMS, openConnection, READ\_PHONE\_STATE, and getNetworkOperator}) for a sample of Opfake, and generates the corresponding semantics (i.e., ``Send SMS messages'', ``Access the Internet'', and ``Collect IMEI'') and malicious behavior description (i.e., ``Send SMS to premium-rate numbers, collect info on the device, and send it to a remote server over the internet''). The malicious behavior description generated by \mytool is also consistent with the expert analysis of Opfake shown in Table \ref{tb5}, which also accurately explains why the sample in Opfake is classified as malware.

In addition to the two examples above, the malicious behavior descriptions of the other samples also match the expert analysis as shown in Table \ref{tb5}. \mytool provides a fairly reasonable explanation for the classification results. However, there are also some exceptions, such as a sample of FakeInstaller shown in Table \ref{tb5}. The malicious behavior description generated by \mytool includes the behavior of sending SMS to premium-rate numbers, collecting information on the device, and keeping running in the background, however, the expert analysis only includes the behavior of sending the premium SMS. After manual analysis, we find that this sample indeed has the behavior of collecting information and keeping running in the background. Another sample is ``xunbaikew1'', which collects contact information and sends SMS message with the app download link to all contacts. \mytool captures the malicious behavior of collecting contact information but 
{misses} the behavior of sending SMS message as sending SMS to a premium-rate number. Actually, some key features such as SEND\_SMS can be mapped to different malicious behaviors in different scenarios like sending SMS with malicious download links. \mytool may not be able to cover all the malicious behaviors only by mapping the key features. It can be improved by adding more other information from apps. Based on the expert analysis of the top 16 malware families, we find that 13 malware families (except FakeInstaller, FakeDoc, and SendPay) have the behavior of stealing information and sending it to a remote server over the internet, and 7 malware families have the behavior of sending SMS messages. Moreover, some of the information stolen by malware families is the same (e.g., IMEI, OS version, and device ID). {We can conclude that the APIs and permissions used to perform malicious behaviors between different malware families are similar in Drebin dataset,} which is consistent with our experimental results. 

As aforementioned in Section~\ref{subsec:accuracy}, two benign apps are misclassified as malware and three malware samples are misclassified as benign. We attempt to analyze why they are misclassified according to the interpretable results of \mytool. 
The two benign applications that are misclassified are HiViewTunnel and HwSpaceService, which are internal system applications for the HUAWEI phone. We can see in Table \ref{tb6} that the two apps do use some suspicious permissions and APIs, causing them being classified as malware. In fact, they are just built-in system apps that use sensitive APIs and permissions.
In this case, it is difficult for \mytool to correctly distinguish malware, as the built-in system apps have the same features and behaviors as malware. 
For the three malware samples that are misclassified, \mytool outputs no key features for all of them, 
which means that \mytool does not identify any key features of these samples to classify them as malware, resulting in malicious samples being misclassified as benign. We further manually analyze these three samples and find that the malware APK file in SMSreg lacks the configuration file, AndroidManifest.xml, resulting in that the app has no permission to perform malicious behaviors so as to be identified as benign. 
The remaining two samples hide malicious behavior in the .so file and the asset folder, causing their malicious behaviors to be unrecognizable because \mytool does not analyze the .so files and the files in the asset folder.

\subsubsection{\revised{Online Survey}}
\revised{In order to alleviate the bias caused by subjective opinions, we conduct an online survey to investigate the respondents' evaluation of the interpretable results (i.e., malware description) generated by \mytool.}

\vspace{1mm}
\noindent{\textbf{Dataset}.}
\revised{We randomly select one sample from each malware family and use the corresponding interpretability results of these malware to design the survey.}

\begin{figure}\centering
\includegraphics[width=0.55\textwidth]{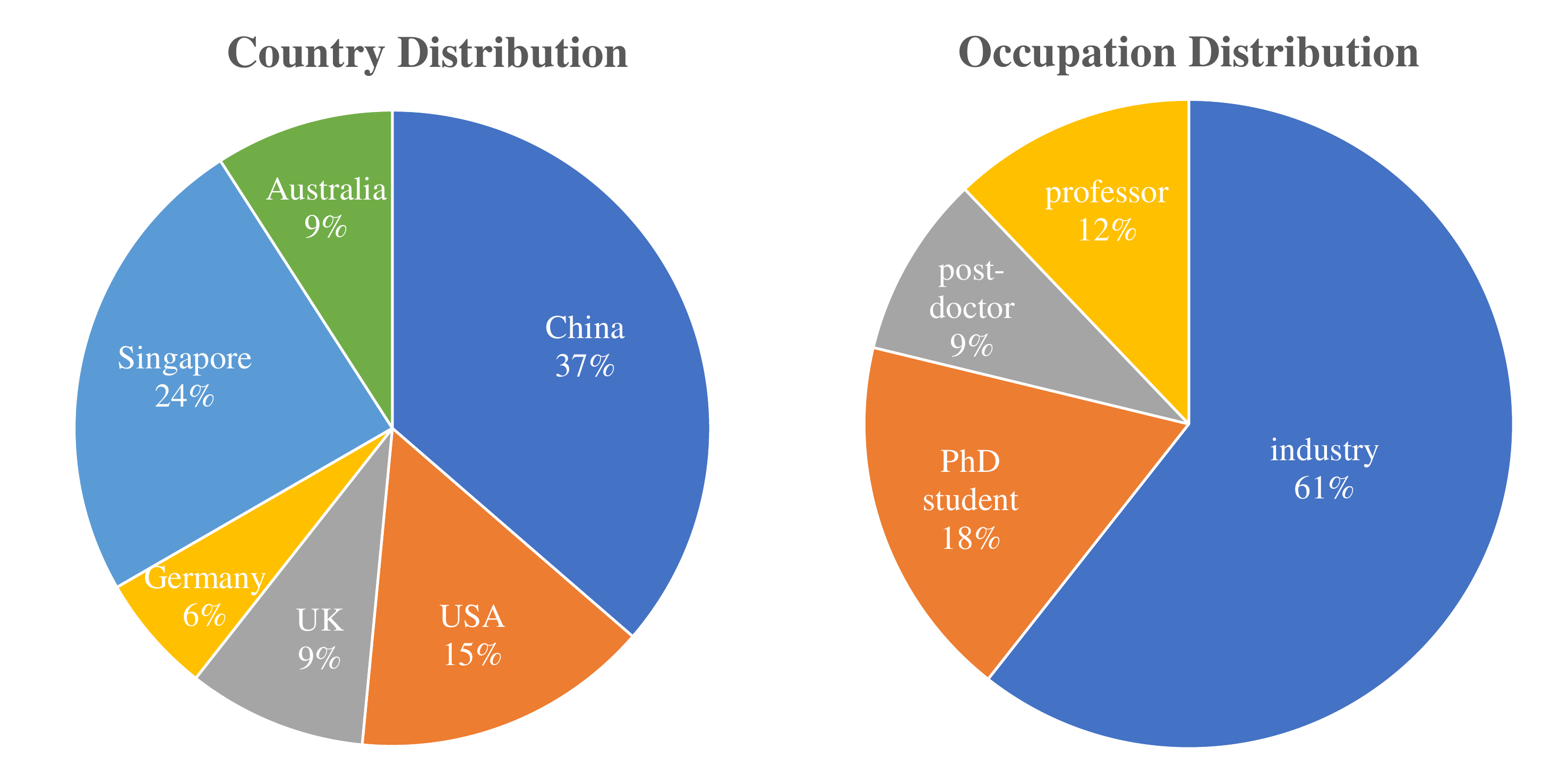}
\caption{\revised{Country and occupation distribution of participants}}
\label{fig:distribution}
\end{figure}

\vspace{1mm}
\noindent{\textbf{Participant Recruitment.}}
\revised{We recruit {33} people from industrial companies and our universities to participate in the experiments via emails and word-of-mouth. Among the participants, \revised{60.6\%} come from industry, and the rest come from academia. Note that, 6 security analysts from the National Internet Emergency Center also corroborate with us and help to accomplish this online survey. They come from different countries, such as USA, UK, Germany, China, Singapore, and Australia. They have a variety of occupations, ranging from \revised{PhD} students, post-doctoral researchers, and professors. \revised{Fig.~\ref{fig:distribution} shows the country and occupation distribution of participants}. {Their expertise includes} app developers, computer security {professionals}, and machine learning researchers. Among them, 20 participants have experience in malware classification, while 17 respondents have more than 1 year of Android malware classification experience.}

\vspace{1mm}
\noindent{\textbf{Experiment Procedures.}}
\revised{We start the online survey with a brief introduction. We explain to the participants that our task is to evaluate how well the generated malware description results match the ground truth. Then the participants are required to provide their personal information relevant to the survey. In order to quantitatively measure the quality of the generated results, we define the rating scale as 1 to 5 where a higher score means that the generated results match the ground truth better. The participants are required to rate the quality of the generated results by comparing {them} with the ground truth of a total of 16 malware samples from 16 malware families.}

\revised{There are two main tasks that participants are required to complete. Participants need to 
(1) fill in their personal information in the survey, such as name, country, academia or industry, field of work, the role at work, and their experience in Android malware classification, etc.{, and to} (2) click the corresponding button in the survey to rate the generated malware descriptions of each sample by comparing {them} with the ground truth.
The online survey contains 26 questions in total and takes about 20 minutes to complete. Table~\ref{tb17} demonstrates a part of the questions in the survey.} \revised{The survey is available on ~{\url{https://forms.gle/RFUmPaSE9eKfG9zm8}}.}


\begin{table}[b]\footnotesize
\centering
\caption{\revised{Part of the questions in the survey}}
\scalebox{0.9}{
\begin{tabular}{l|l|l}
\hline
\textbf{ } & \textbf{Questions} & \textbf{Rating Scale (Score: 1-5)} \\\hline
\multirow{9}{*}{Part2} & \multirow{5}{*}{\begin{tabular}[c]{@{}l@{}}Q1: score the following generated description of \qq{Adrd0}. \\ 
\qquad \textbf{Ground Truth:}\\
\qquad	1. Activate when the mobile device is booted up.\\
\qquad    2. Access the Internet and download componets\\
\qquad    3. Steal the following info and send to a remote server.\\
\qquad \textbf{Generated Result:}\\
\qquad	Launch with system startup, collect info on the device\\
\qquad  and send  it to remote server over the internet.\\
Q2: score the following generated description of \qq{BaseBridge0}\\
\qquad ......
\\\end{tabular}} & \multirow{5}{*}{\begin{tabular}[c]{@{}l@{}}1: Poor\\ 2: Marginal\\ 3: Acceptable\\ 4: Good\\ 5: Excellent \end{tabular}} \\
 &  &  \\
 &  &  \\
 &  &  \\
 &  &  \\
 &  &  \\
 &  &  \\
  &  &  \\
 &  &  \\
 &  &  \\ \hline
\end{tabular}
}
\label{tb17}
\end{table}

\begin{figure}
\begin{minipage}[t]{0.475\linewidth}
\centering
\includegraphics[width=2.7in]{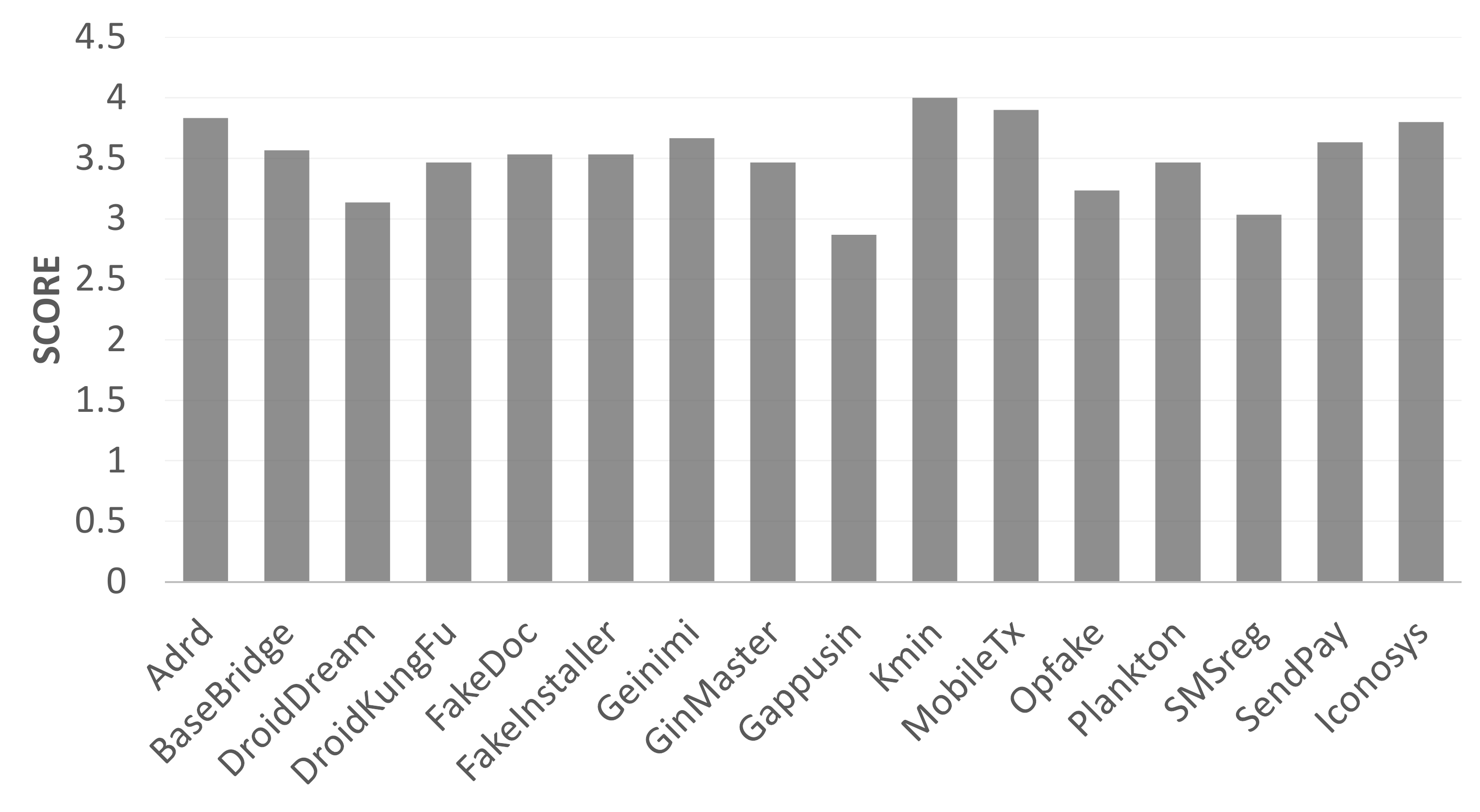}
\caption{\revised{The average score of all samples from survey}}
\label{survey}
\end{minipage}%
\begin{minipage}[t]{0.475\linewidth}
\centering
\includegraphics[width=2.7in]{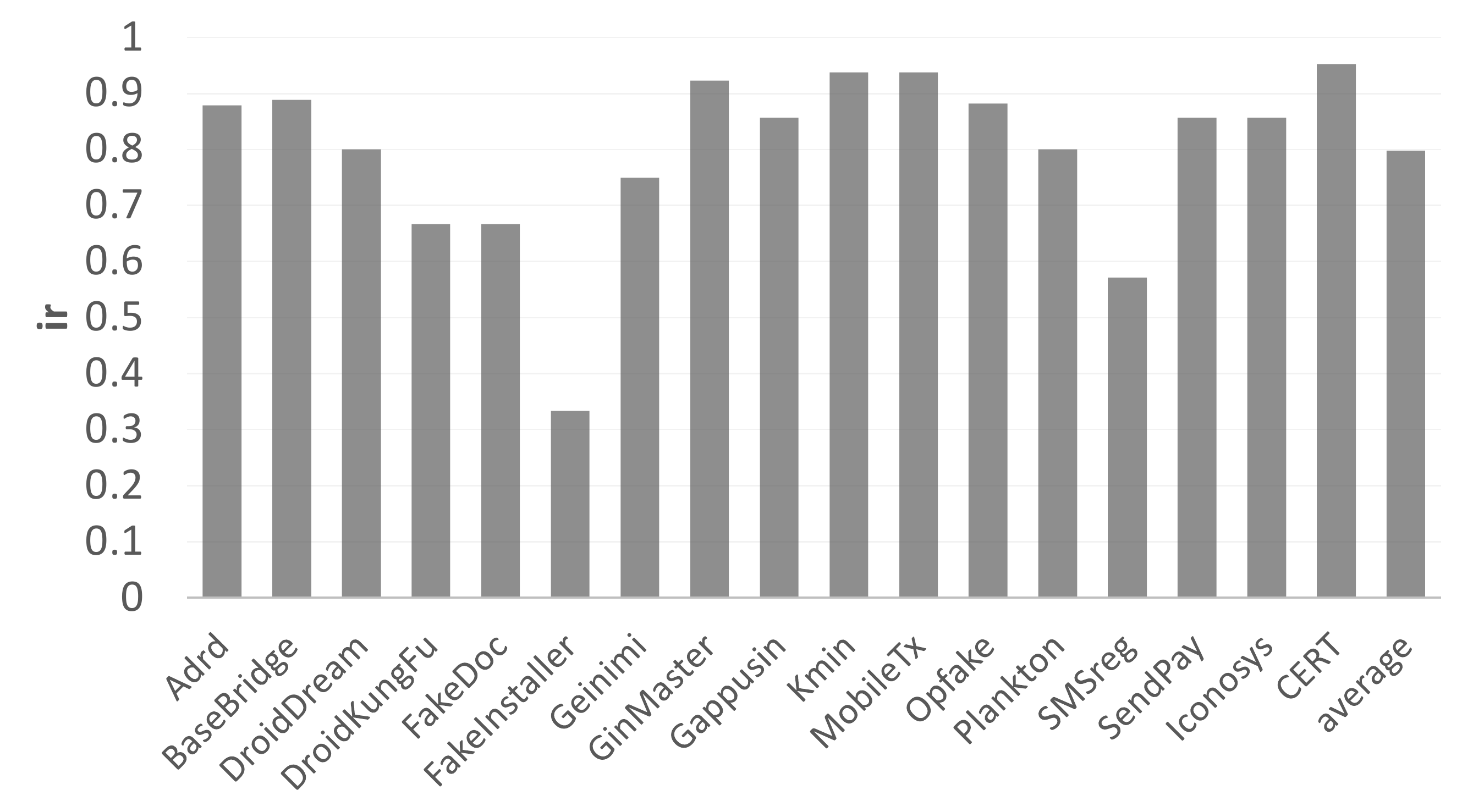}
\caption{\revised{The \textit{ir} results of all family samples}}
\label{rq2}
\end{minipage}
\end{figure}

\vspace{1mm}
\noindent{\textbf{Survey Results}.}
\revised{
To ensure the quality of the survey result, we excluded those surveys that take less than 5 minutes to complete, and finally obtained 30 valid survey results.
The average score of each sample is shown in Fig.~\ref{survey}. The average score of Kmin is 4, which means that the generated result of Kmin is good. Except for Gappusin, all other sampled scores are more than 3, which means that the generated description results are acceptable. All in all, the average score of all samples is \revised{3.7}. Therefore, we can conclude that the overall generated description result is {better than Acceptable and close to Good.}
}


\subsubsection{Quantitative Analysis}
\revised{In the end, besides the human evaluation, we further conduct a quantitative study to investigate how well the generated description matches the ground truth of the malware.}
\revised{Here, we define the 10 samples from the National Internet Emergency Center as the CERT family. The \textit{ir} of interpretability results across all malware samples generated by \mytool is shown in Fig.~\ref{rq2}. We can see that the average \textit{ir} of all malware samples is {0.80}, while the \textit{ir} of four families (i.e., GinMaster, Kmin, MobileTx, and CERT) is greater than 0.9.
The \textit {ir} of FakeInstaller and SMSreg families is lower than others, only 0.33 and 0.58 respectively. \revised{The reason is that} most samples in FakeInstaller and SMSreg malware families have no more than two malicious behaviors, but \mytool utilizes 6 key features to generate the malware description, resulting in a lot of \textit{surplus\_concepts} in these two malware families. To summarize, \mytool achieves good interpretability results for all malware families except for FakeInstaller and SMSreg \revised{whose samples only have no more than 2 malicious behaviours}.} 


\subsection{\revised{RQ3: Does \mytool achieve a better interpretation than the state-of-the-art techniques?}} 

\revised{In this experiment, we aim to demonstrate that \mytool can get a better interpretation than the state-of-the-art techniques. To achieve this goal, we conduct experiments on \mytool, Drebin and LIME under the same data set, and compare their interpretation results through quantitative analysis and case 
studies.}

\subsubsection{Setup.}
\revised{In this experiment, to demonstrate the interpreting effectiveness of our \mytool, we compare it with two state-of-the-art interpretable ML systems, Drebin~\citep{arp2014drebin} and LIME~\citep{ribeiro2016should}, on the 170 malicious samples selected and used in RQ2. 
The reasons we choose Drebin and LIME have been mentioned in Section ~\ref{subsec:procedure1}. For Drebin, we obtain the weight of features by acquiring the coefficient of the liner SVM. The features with the highest weights are regarded as key features and used to interpret the classification result. Similarly, we perform LIME on the MLP model to obtain the feature weight of each test sample, and then select the feature with the highest weight as the key feature and use it to explain why an app is classified as malware or benign app.}

\revised{In order to compare the three methods, we first calculate all key features for all samples and choose 6 key features in this comparison experiment. Note that the number of key features generated by LIME is 6 by default. For fairness, the number of key features generated by Drebin is also set to 6. We then obtain the corresponding semantics of the key features based on the semantic database, which allows us to understand the connection between the key features and the ground truth. Finally, we compare the interpretability of these three methods based on how well the key features and semantics match the expert analysis, and discuss the performance of the three methods based on the key features of the three malware families generated by them.} 

\revised{In addition, we compare the three methods from a quantitative perspective. We extract the \qq{concepts} from the semantics of key features, and compute the \textit{detect\_concepts} and \textit{surplus\_concepts}. \textit{total\_concepts} can be extracted from the ground truth of malware samples. Then we respectively calculate the \textit{ir} of the interpretation results generated by the three methods and make a comparison. {Moreover, we compare the total number of key features across all these malware samples generated by the Drebin, LIME, and \mytool.}}

\subsubsection{Results.}

\revised{We compare the three methods on the 170 malware samples, and demonstrate the results of five representative malware (i.e., Adrd, GinMaster, MobileTx, blackgame, and xunbaikew1), as shown in Table~~\ref{tb7}.} For the sample in {Adrd}, \mytool outputs the key features, among which {READ\_PHONE\_STATE}, {LocationManager.request} and {getSubscriberId} match the behavior of collecting confidential information (e.g., location and IMEI/IMSI), and the features, {openConnection} and {getResponseCode}, match the behavior of sending information to remote location over the internet, the remaining feature, {RECEIVE\_BOOT\_COMPLETED}, matches the behavior of launching with system startup. By contrast, Drebin outputs the key features, among which {getContent}, {getDeviceId}, and {openConnection} match the behavior of stealing information and sending it to a remote location. However, the other features (i.e., {Intent.action.MAIN}, {INSTALL\_PACKAGES} and NotificationManager.cancel) do not match the key malicious behaviors. The key features generated by LIME include openConnection, RECEIVE\_BOOT\_COMPLETED and getDeviceId, which can match the behavior of re-executing itself when the device is booted up and collecting information. But the remaining key features, like {INSTALL\_PACKAGES} and ContentResolver.delete cannot match any behaviors of Adrd.

{Similarly, for blackgame, xunbaikew1, and the samples in GinMaster and MobileTx,} \mytool outputs key features that match the malicious behavior of the corresponding {sample. } However, some of the key features generated by {LIME and Drebin cannot match the behaviors of {blackgame, xunbaikew1,} GinMaster and MobileTx, as shown in the bold features in Table \ref{tb7}}. For the sample in GinMaster, the key features generated by LIME can match most of the behavior of GinMaster, but the remaining features (i.e., {NotificationManager.notify} and {NotificationManager.cancel}) do not match any behaviors. For the sample in MobileTx, Drebin generates some key features that match the malicious behavior of stealing information and sending SMS messages to a premium-rate number, but the remaining features {(e.g, {Intent.action.MAIN} and {INSTALL\_PACKAGES})} do not match any malicious behavior of MobileTx. Blackgame and xunbaikew1 have similar phenomena with GinMaster and MobileTx.

\begin{table*}\scriptsize
\centering
\caption{\revised{Comparison of three approaches (i.e., Drebin, LIME, and XMal). {The bold texts refer to key features that cannot match the real malicious behavior. Ground Truth refers to the expert analysis corresponding to each sample.}}}
\label{tb7}
\scalebox{0.8}{
\begin{tabular}{c|l|l|l}
\hline
\multirow{7}{*}{\multirow{7}{*}{\rotatebox{90}{Adrd}}} & \multicolumn{1}{c|}{Drebin} & \multicolumn{1}{c|}{LIME} & \multicolumn{1}{c}{XMal} \\ \cline{2-4} 
 & \begin{tabular}[c]{@{}l@{}}\textcolor{black}{\cellcolor{gray!30}\textit{Key Features:}}\\ {\textbf{Intent.action.MAIN}}\\\textbf{INSTALL\_PACKAGES}\\URL.getContent\\ TelephonyManager.getDeviceId\\ URL.openConnection\\ \textbf{NotificationManager.cancel}\\{\cellcolor{gray!30}\textit{Corresponding Semantics:}}\\1.None \qquad  2.Install packege\\ 3.Get data from the Internet \ 4.Collect device ID\\ 5.Access the Internet \quad 6.Cancel notification\\\end{tabular}  
 &  \begin{tabular}[c]{@{}l@{}}\textcolor{black}{\cellcolor{gray!30}\textit{Key Features:}}\\{SEND\_SMS}\\
    URL.openConnection\\ 
    RECEIVE\_BOOT\_COMPLETED\\ 
    TelephonyManager.getDeviceId\\ \textbf{INSTALL\_PACKAGES}\\ \textbf{ContentResolver.delete}\\{\cellcolor{gray!30}\textit{Corresponding Semantics:}}\\1.Send SMS message \qquad 2.Access the Internet\\3.Activited by BOOT \qquad 4.Collect device ID(IMEI)\\5.Install packege \qquad \qquad 6.Delete URI data\\  \end{tabular} 
 & \begin{tabular}[c]{@{}l@{}}{\cellcolor{gray!30}\textit{Key Features:}}\\URL.openConnection\\ READ\_PHONE\_STATE\\ RECEIVE\_BOOT\_COMPLETED\\ LocationManager.request\\ HttpURLConnection.getResponseCode\\ TelephonyManager.getSubscriberId\\{\cellcolor{gray!30}\textit{Corresponding Semantics:}}\\1.Access the Internet \qquad 2.Collect phone status\\3.Activited by BOOT \qquad 4.Get updated location\\5.Get Http response code \ \ 6.Collect SubscriberId ID(IMSI)\\
                           \end{tabular}
 \\ \cline{2-4}
 & \multicolumn{3}{l}{\begin{tabular}[c]{@{}l@{}} {\textbf{\textit{\textit{Ground Truth:}}}{1. Activate when the mobile device is booted up}. {2. Access the Internet and download components.} 3. Steal some info and send to remote server}\end{tabular}} \\ \hline
\multirow{7}{*}{\multirow{7}{*}{\rotatebox{90}{GinMaster}}} & \multicolumn{1}{c|}{Drebin} & \multicolumn{1}{c|}{LIME} & \multicolumn{1}{c}{XMal} \\ \cline{2-4} 
 & \begin{tabular}[c]{@{}l@{}}{\cellcolor{gray!30}\textit{Key Features:}}\\\textbf{Intent.action.MAIN}\\TelephonyManager.getDeviceId\\TelephonyManager.getSimSerialNumber\\URL.openConnection\\\textbf{NotificationManager.cancel}\\RECEIVE\_BOOT\_COMPLETED\\{\cellcolor{gray!30}\textit{Corresponding Semantics:}}\\1.None \qquad \qquad 2.Collect device ID(IMEI) \\3.Collect ICCID \qquad \ \qquad 4.Access the Internet\\5.Cancel notification \qquad 6.Activited by BOOT\\
\end{tabular}   
 
 & \begin{tabular}[c]{@{}l@{}}{\cellcolor{gray!30}\textit{Key Features:}}\\RECEIVE\_BOOT\_COMPLETED\\TelephonyManager.getDeviceId\\ \textbf{NotificationManager.notify}\\
                           URL.openConnection\\ TelephonyManager.getSimSerialNumber\\                     \textbf{NotificationManager.cancel}\\{\cellcolor{gray!30}\textit{Corresponding Semantics:}}\\1.Activited by BOOT \qquad 2.Collect device ID(IMEI)\\3.Post notification\qquad 4.Access the Internet\\5.Collect ICCID \qquad 6.Cancel notification\\  \end{tabular}
 
 & \begin{tabular}[c]{@{}l@{}}{\cellcolor{gray!30}\textit{Key Features:}}\\URL.openConnection\\ READ\_PHONE\_STATE\\ RECEIVE\_BOOT\_COMPLETED\\ HttpURLConnection.getResponseCode\\ TelephonyManager.getSubscriberId\\ \\{\cellcolor{gray!30}\textit{Corresponding Semantics:}}\\1.Access the Internet \qquad 2.Collect phone status\\3.Activited by BOOT \qquad \ \ \qquad 4.Get Http response code \\5.Collect SubscriberId ID(IMSI)\end{tabular} 
 
 \\ \cline{2-4}
  & \multicolumn{3}{l}{\begin{tabular}[c]{@{}l@{}}\textbf{\textit{Ground Truth:}} 1. Steal info from the device. 2. Send info to remote server. 3. The malicious service is triggered when the device finishes a boot.\end{tabular}} \\ \hline
 
\multirow{7}{*}{\multirow{7}{*}{\rotatebox{90}{MobileTx}}} & \multicolumn{1}{c|}{Drebin} & \multicolumn{1}{c|}{LIME} & \multicolumn{1}{c}{XMal} \\ \cline{2-4} 
 & \begin{tabular}[c]{@{}l@{}}{\cellcolor{gray!30}\textit{Key Features:}}\\\textbf{Intent.action.MAIN}\\ \textcolor{black}{INSTALL\_PACKAGES}\\URL.openConnection\\ RECEIVE\_SMS\\ \textbf{ActivityManager.restartPackage}\\ SEND\_SMS\\{\cellcolor{gray!30}\textit{Corresponding Semantics:}}\\1.None \ \  \ \ \ 2.Install package\\3.Get data from the Internet \qquad  4.collect SMS\\5.Break other applications \qquad \quad 6.Collect SMS\\\end{tabular} 
 
 & \begin{tabular}[c]{@{}l@{}}{\cellcolor{gray!30}\textit{Key Features:}}\\SEND\_SMS\\TelephonyManager.getDeviceId\\
                           RECEIVE\_SMS\\ \textbf{INSTALL\_PACKAGES}\\
                           READ\_SMS\\    \textbf{ActivityManager.restartPackage}\\{\cellcolor{gray!30}\textit{Corresponding Semantics:}}\\1.Send SMS message \qquad 2.Collect device ID(IMEI)\\3.Collect SMS \qquad 4.Install package\\5.Collect SMS \quad 6.Break other applications\\
                \end{tabular} 
 
 & \begin{tabular}[c]{@{}l@{}}{\cellcolor{gray!30}\textit{Key Features:}}\\SEND\_SMS\\ URL.openConnection\\ READ\_PHONE\_STATE\\ TelephonyManager.getSubscriberId\\ HttpURLConnection.getResponseCode \\ \\{\cellcolor{gray!30}\textit{Corresponding Semantics:}}\\1.collect SMS \qquad 2.Access the Internet\\3.Collect phone status \quad 4.Collect SubscriberId ID(IMSI) \ \ \ \\5.Get Http response code \\\end{tabular} 
 \\ \cline{2-4}
  & \multicolumn{3}{l}{\begin{tabular}[c]{@{}l@{}}\textbf{\textit{\textit{Ground Truth:}}} 1. Steal info from the compromised device.  2. Send SMS messages to premium-rate number.\end{tabular}} \\ \hline
  
\multirow{7}{*}{\multirow{7}{*}{\rotatebox{90}{blackgame}}} & \multicolumn{1}{c|}{Drebin} & \multicolumn{1}{c|}{LIME} & \multicolumn{1}{c}{XMal} \\ \cline{2-4} 
 & \begin{tabular}[c]{@{}l@{}}{\cellcolor{gray!30}\textit{Key Features:}}\\\textbf{Intent.action.MAIN}\\ \textcolor{black}{WifiManager.setWifiEnabled}\\TelephonyManager.getDeviceId\\TelephonyManager.getSimSerialNumber\\URL.openConnection\\ \textbf{NotificationManager.cancel}\\{\cellcolor{gray!30}\textit{Corresponding Semantics:}}\\1.None \ \ \  2.Check whether  wifi is enabled\\3.Collect \ device \ ID(IMEI) \ \ \ 4.Collect \  ICCID\\5.Access the Internet \ \ \ \ \ \ \ \ \ 6.Cancel notification\\\end{tabular} 
 
 & \begin{tabular}[c]{@{}l@{}}{\cellcolor{gray!30}\textit{Key Features:}}\\SmsManager.sendDataMessage\\WifiManager.setWifiEnabled\\
                           \textbf{RECEIVE\_MMS}\\\textbf{ContentResolver.delete}\\TelephonyManager.getNetworkOperatorName\\elephonyManager.getDeviceId\\{\cellcolor{gray!30}\textit{Corresponding Semantics:}}\\1.Send SMS message \ 2.Check whether wifi enabled \\3.Collect MMS \quad 5.Collect network operator name \\4.Delete URI data \quad  6.Collect device ID(IMEI)\\
                \end{tabular} 
 
 & \begin{tabular}[c]{@{}l@{}}{\cellcolor{gray!30}\textit{Key Features:}}\\URL.openConnection\\ SEND\_SMS\\RECEIVE\_SMS\\WRITE\_SMS\\TelephonyManager.getDeviceId\\TelephonyManager.getSubscriberId\\{\cellcolor{gray!30}\textit{Corresponding Semantics:}}\\1.Access the Internet \qquad 2.Send SMS message \\3.Collect SMS \qquad 4.Write SMS \\5.collect device ID(IMEI) \ \ \ 6.collect SubscriberId ID(IMSI)\\
\end{tabular} 
 \\ \cline{2-4}
  & \multicolumn{3}{l}{\begin{tabular}[c]{@{}l@{}}\textbf{\textit{\textit{Ground Truth:}}} 1. Send SMS message to premium-rate num. 2. Obtain phone num and device info and upload it to the remote server.\end{tabular}} \\ \hline
  
\multirow{7}{*}{\multirow{7}{*}{\rotatebox{90}{xunbaikew1}}} & \multicolumn{1}{c|}{Drebin} & \multicolumn{1}{c|}{LIME} & \multicolumn{1}{c}{XMal} \\ \cline{2-4} 
 & \begin{tabular}[c]{@{}l@{}}{\cellcolor{gray!30}\textit{Key Features:}}\\\textbf{Intent.action.MAIN}\\ \textcolor{black}{ContentResolver.query}\\SEND\_SMS\\\textbf{Runtime.exec}\\READ\_CONTACTS\\\textbf{PowerManager.newWakeLock}\\{\cellcolor{gray!30}\textit{Corresponding Semantics:}}\\1.None \ 2.Query URL data \ 3.Send SMS message \\4.Execute command \quad 5.Collect contacts\\ 6.Keep processor and screen awake\\\end{tabular} 
 
 & \begin{tabular}[c]{@{}l@{}}{\cellcolor{gray!30}\textit{Key Features:}}\\SEND\_SMS\\ContentResolver.query\\
                           RECEIVE\_SMS\\ \textbf{INSTALL\_PACKAGES}\\
                           READ\_SMS\\    \textbf{ActivityManager.restartPackage}\\{\cellcolor{gray!30}\textit{Corresponding Semantics:}}\\1.Send SMS message \qquad \qquad 2.Query URL data \\3.Collect SMS \qquad \qquad 4.Install package\\5.Collect SMS \qquad \qquad \ \  6.Break other applications\\                \end{tabular} 
 
 & \begin{tabular}[c]{@{}l@{}}{\cellcolor{gray!30}\textit{Key Features:}}\\SEND\_SMS\\ ContentResolver.query\\READ\_CONTACTS\\ \\ \\ \\{\cellcolor{gray!30}\textit{Corresponding Semantics:}}\\1.Send SMS message \qquad \qquad \qquad  \ \ \ \quad 2.Query URL data \\ 3.Collect contacts\\ \\ \end{tabular} 
 \\ \cline{2-4}
  & \multicolumn{3}{l}{\begin{tabular}[c]{@{}l@{}}\textbf{\textit{\textit{Ground Truth:}}} Collect contact info, and then send SMS with the app download link to all contacts.\end{tabular}} \\ \hline  
\end{tabular}
}
\end{table*}

\begin{figure*}
\centering
\includegraphics[width=0.8\textwidth]{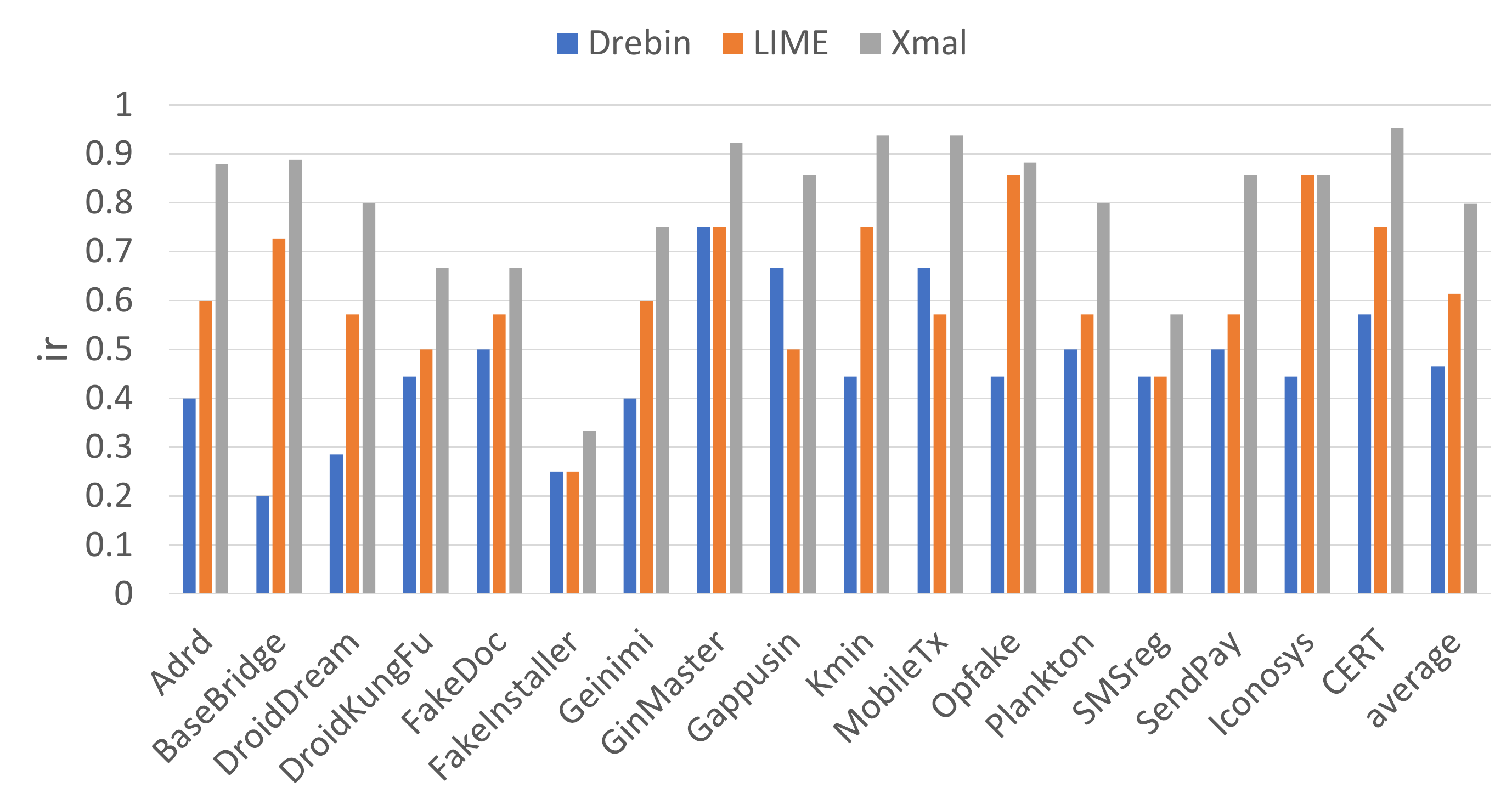}
        \caption{\revised{The \textit{ir} results across all malware samples generated by Drebin, LIME, and \mytool}}
\label{rq3}
\end{figure*}

For Drebin, the feature with maximum weight in Adrd, GinMaster, MobileTx, {blackgame, and xunbaikew1} is always {Intent.action.MAIN}, and some key features can not reveal any malicious behaviors.
The reasons are as follows. (1) Drebin utilizes the simple detection of linear SVM to determine the contribution of each individual feature to the classification and the feature weight of the model is only related to the model, but not to the test sample. {If the features exist in test sample and the features have a large weight in the model, they will be selected as key features.} Therefore, it makes sense that why {Intent.action.MAIN} is always the key feature and some key features generated by Drebin can not reveal the malicious behaviors. For LIME, as we can see, the key features generated by it do not match the behaviors of malware families very well. For example, LIME outputs the feature with maximum weight, i.e., {NotificationManager.notify} and {NotificationManager.cancel}, which do not match any malicious behaviors of GinMaster. The key features generated by LIME may not be accurate enough to give a reasonable explanation of the classification result in Android malware detection. (2) LIME generates a linear model to approximate the local part of the original complex model, which makes it difficult for LIME to accurately approximate the decision boundaries near an instance, especially in malware detection applications. For \mytool, it generates key features that closely match the behaviors of the malware families.

\begin{figure*}
\centering
\includegraphics[width=1\textwidth]{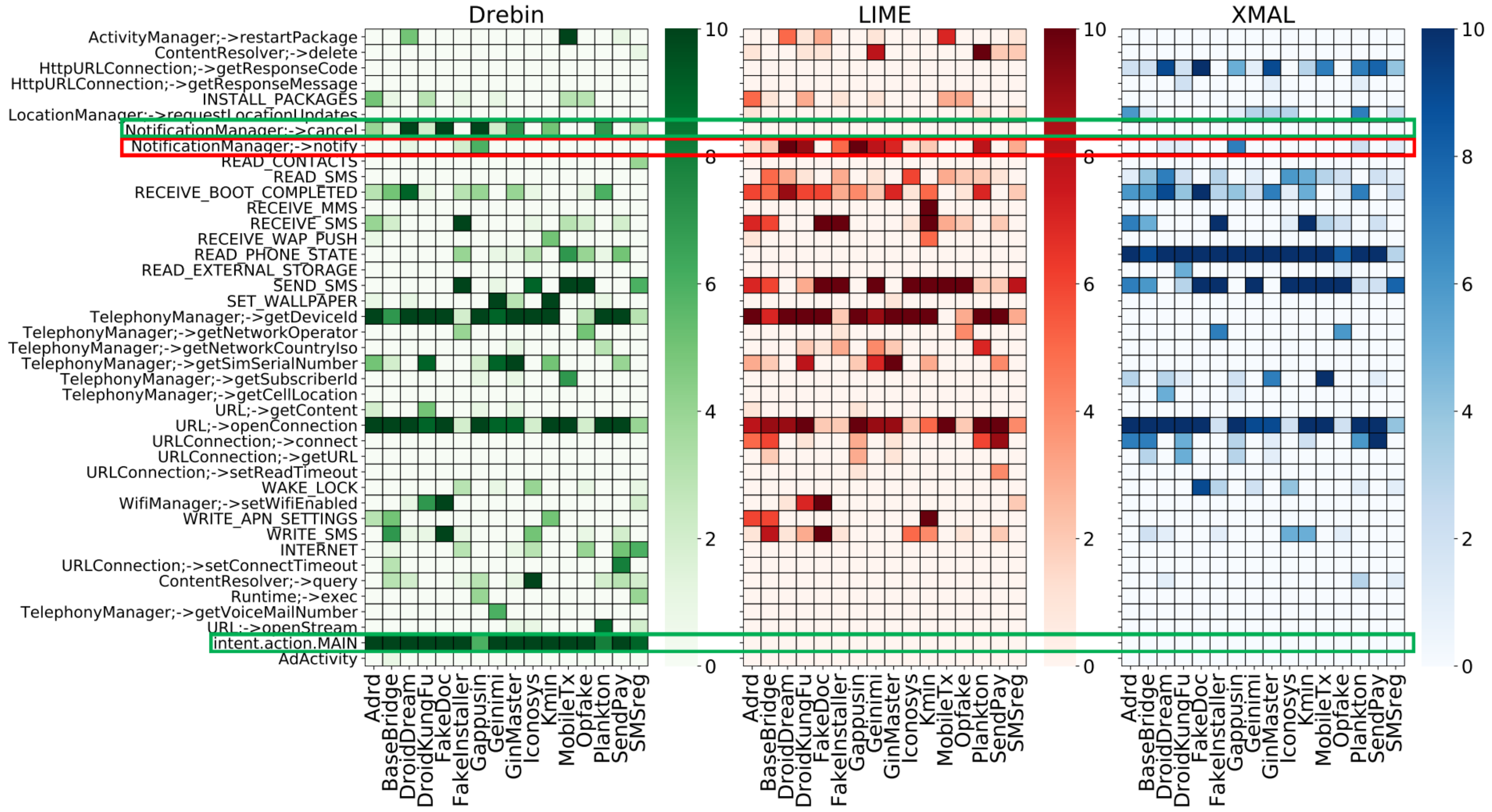}
        \caption{{Total number of key features across all malware samples generated by the Drebin, LIME, and \mytool}}
\label{fig3}
\end{figure*}

\revised{We also conduct a quantitative analysis for the three methods. The \textit{ir} of the interpretability results across all these malware samples generated by Drebin, LIME and \mytool is shown in Fig.~\ref{rq3}. \mytool achieves the best interpretability results among the three methods across all these malware families. LIME is better than Drebin in most malware families except for \textit{Gappusin} and \textit{MobileTx}. \mytool obtains the largest average value of \textit{ir} across all malicious families.}
The total number of the key features across all malware samples generated by the three method is shown in Fig.~\ref{fig3}. Specifically, (1) openConnection is a common key feature for all families generated by the three methods, which indicates that most of malicious behaviors are based on Internet for these families.
(2) Drebin and LIME output the same common feature TelephonyManager.getDeviceId for all families. This feature is used to get mobile information (e.g., IMEI). \mytool outputs {READ\_PHONE\_STATE} with the similar function as getDeviceId. (3) Similarly, {SEND\_SMS} is another common key feature for LIME and \mytool, however, Drebin cannot identify {SEDN\_SMS} for some malware families such as BaseBridge and {Kmin}. Both of them contain the behavior of sending SMS. The feature {RECEIVE\_BOOT\_COMPLETED} generated by LIME and \mytool has the similar phenomenon with SEND\_SMS. (4) Drebin generates two other common key features (i.e., {Intent.action.MAIN} and {NotificationManager.cancel}) for most families, but both of them cannot reveal malicious behaviors, \revised{as shown in Fig.~\ref{fig3} (green box)}.
The similar phenomenon occurs on LIME for the feature {NotificationManager.notify}, \revised{as shown in Fig.~\ref{fig3} (red box)}.

In summary, Drebin generated some key features that cannot reveal malicious behaviors such as {Intent.action.MAIN}. LIME has a better performance in these families, but sometimes generates some key features that are  meaningless to interpret the malicious behaviors in concrete cases (shown in Table~\ref{tb7}). \mytool generates key features for most malware families and is able to reveal the key malicious behaviors within apps. Therefore, \mytool achieves a better performance on interpretability of Android malware detection.

\section{Discussion}
In this section, we discuss the threats to validity, limitations of \mytool and summarize open challenges in the interpretability of Android malware detection according to our study.

\subsection{Threats to validity}

\revised{\textbf{Conclusion validity. }}\revised{Since we use the collected ground truth to validate the malware descriptions generated by our method, the results of \mytool may differ if the ground truth is not accurate itself. To ensure the accuracy of the ground truth, we collaborate with an experienced expert team from National Internet Emergency Center and they manually analyze malware samples and provide the analysis reports. We also collect the corresponding expert analysis reports of each family from Symantec and Microsoft, and cross-validate the analysis reports from the two different resources and obtain the final ground truth of malware descriptions.}

\noindent \revised{\textbf{Construct validity. }}\revised{We evaluate the generated results by manual comparison with the ground truth, the results may be biased by our subjective opinions. In order to mitigate this issue, we randomly select one sample from each malware family and conduct an online survey to investigate the quality of the generated malware description for these samples.}

\noindent \revised{\textbf{Internal validity. }}\revised{The performance of classification models depends on {the} training set. If the training set is small and not representative, the model cannot achieve a good generalization ability. We collect {15,570} 
{malicious samples}, including all varieties of threats for Android, such as data leakage, phishing, trojans, spyware, and root exploits. These malware samples are the recent malicious samples rather than the old dataset such as Genome~\citep{zhou2012dissecting} in 2011. Moreover, the performance of our method also depends on the hyperparameter configuration. It is important to select a proper value for \textit{n}. We perform hyperparameter tuning and explain the procedures in Section ~\ref{subsec:procedure2}. 
}

\noindent \revised{\textbf{External validity.} }\revised{Additionally, we simplify and generalize the functional descriptions into simple semantics by intercepting and generalizing the key predicates, objects, and complements. Therefore, semantics constitute is a threat to the external validity of the experiments. 
{For reproducibility purpose}, we release the functional 
{descriptions} and the constructed semantics on our website~{\url{https://sites.google.com/view/xmal/}}.}

\subsection{Limitations of \mytool}
We have proven the good performance of \mytool in classification accuracy and interpretability, but it still has some limitations as follows.
A malware sample can contain several different malicious behaviors. For example, a sample in the family, Geinimi, may collect information (e.g., IMEI, location, SMS messages, and contact) and upload them to a remote server, send SMS messages to a premium-rate number, install or uninstall software and create a shortcut. \mytool cannot output features that match all the malicious behaviors in the sample, because it makes predictions by focusing on the features with the highest weights (e.g., sending SMS messages to premium-rate number), causing it to only notice the malicious behavior of a certain part. It may be possible to improve it by using a multi-attention mechanism, which is our future work. To mitigate this issue, it may be helpful to use multi-attention to focus on different parts of features or different types of features. Multi-attention~\citep{kim2018multimodal} utilizes a multi-modal deep learning method to learn various kinds of features, and uses multiple attentions to focus on more features and behaviors in order to identify more malicious behaviors. 

The work in this paper is to explain why an app is classified as malware based on APIs and permissions. Although these two features can effectively target malicious behaviors and explain the classification result, they are not enough to explain how the entire malicious behaviors are implemented. More features should be taken into consideration. For example, If malware attempts to {activate itself}
when the mobile device is booted up, it first has to register an intent.action.BOOT\_COMPLETED intent-filter and apply for the RECEIVE\_BOOT\_COMPLETED permission in AndroidManifest.xml file, and then wait to receive a RECEIVE\_BOOT\_COMPLETED intent sent by the Android system in order to launch with system startup. In this case, Intent is a key feature to explain how malicious behaviors are implemented. Therefore, Intent should be taken into consideration in the analysis.
However, considering more features might also result in a decrease in interpretability. When we obtain more key features, it might be more difficult to interpret the classification results. 
{Therefore, it is also an important task to select reasonable features and make a trade-off between the number of features and interpretability performance.}

\revised{Moreover, a small number of advanced malware try to hide their malicious behaviors by using native and reflective calls~\citep{garcia2018lightweight}. Our method can only analyze the APIs that access native and reflective calls to determine whether they are malicious. It is difficult to detect their malicious behaviors when the malicious payloads are only introduced by native code. 
Although our method did not use the native and reflective calls as features, nevertheless, it can still achieve a high detection accuracy of 98.37\%. Besides, native and reflective calls do not have the developer documentations and a detailed functional description like permissions and APIs. As a result, we cannot construct their semantics for interpretability purpose. 
We therefore do not take these features into account in the feature set.}

\subsection{Open Challenges in Interpretability of Malware Detection}
Many open challenges exist in explaining why an app is classified as malware. (1) One of them is a complex scenario. Some dangerous APIs and permissions may be used in benign apps for good purposes, such as the internal system apps. It is a great challenge for approaches that are based on features to predict and interpret. Features that are used in different scenarios may have different purposes. For example, in Section~\ref{subsec:accuracy}, two benign apps are misclassified as malware because they have dangerous API calls and permissions and are considered to perform malicious behaviors. But in fact, they are internal system apps, which own similar features and perform similar behaviors, such as monitoring the phone status. 
(2) Another challenge is the malicious behaviors of current malware become more complex. Malware may hide their behaviors through code obfuscation \citep{linn2003obfuscation} and evade malware detection by downloading the payload after installation. For example, samples in fakeInstaller try to avoid analysis through code obfuscation and recompilation. The malware author modifies its DEX file with an obfuscated version of the recompiled code and uses anti-reverse techniques to avoid dynamic analysis and prevent malware from running in the emulator.
Even with manual analysis, it is difficult to fully understand all the malicious behaviors of some complex malware as we need to analyze more code, API calls, permissions, or other features to locate and explain malicious behavior. 
However, it seems that the current interpretable machine learning methods only use a small portion of features to explain the malicious behaviors.
There is still a long way to go to explain why an app is classified as malware for all malicious samples. 

\section{Related Work}

\subsection{Machine Learning-based Android Malware Detection}
Since the traditional malware detection methods cannot handle an increasing number of malicious apps~\citep{tang2019large,chen2019gui}, machine learning methods have become very popular and have achieved great success in Android malware~\citep{aafer2013droidapiminer, yerima2013new,rastogi2013droidchameleon, arp2014drebin, chen2016stormdroid, chen2016towards,wu2016effective,fereidooni2016anastasia,fan2016poster, li2018significant,chen2018automated}. For example, Aafer et al. \citep{aafer2013droidapiminer} proposed to train a KNN classifier by learning relevant features extracted at API level and achieved accuracy as high as 99\% with a false positive rate as low as 2.2\%. Yerima et al. \citep{yerima2013new} presented a method to detect Android malware based on Bayesian Classification models obtained from API calls, system commands and permissions. Wu et al. \citep{wu2016effective} adopted the k-nearest neighbour classification model that leveraged the use of data-flow APIs as classification features to detect Android malware. Li et al. \citep{li2018significant} utilized three levels of pruning by mining the permission data to identify the most significant permissions and trained an SVM classifier with 22 significant permissions. Other machine learning algorithms such as SVM~\citep{arp2014drebin}, Random forest~\citep{rastogi2013droidchameleon}, and XGboost~\citep{fereidooni2016anastasia} were also used to detect malware and have proven to be effective.

With the popularity of deep neural networks, people began to utilize the deep neural network models for malware detection~\citep{yu2014towards,mclaughlin2017deep,kim2018multimodal, xu2018deeprefiner,fengmobidroid}. Yu et al.~\citep{yu2014towards} proposed to train a malware detection model by using a representative machine learning technique, called ANN. McLaughlin et al. \citep{mclaughlin2017deep} proposed a malware detection system that used a deep convolutional neural network to learn the raw opcode sequence from a disassembled program. Kim et al. \citep{kim2018multimodal} utilized a multi-modal deep learning method to learn various kinds of features in order to maximize the benefits of encompassing multiple feature types. Xu et al. \citep{xu2018deeprefiner} used a Long Short Term Memory on the semantic structure of Android bytecode and applied Multi-layer Perceptron on the XML files in order to identify malware efficiently and effectively. 
All these method focused the malware detection accuracy rather than the malware interpretability.

{\subsection{Machine Learning Interpretability}}
People would like to interpret the machine learning models through visualization and behavior interpreting, which is what we are going to introduce.

\subsubsection{Visualization}
Visualization plays an important role in interpreting the machine learning algorithm, especially 
{dimensionality} reduction, clustering, classification and regression analysis. Elzen et al. \citep{van2011baobabview} proposed a system that provided an intuitive visual representation of attribute importance within different levels of the decision tree, helping users to gain a deeper understanding of the decision tree result. Park et al. \citep{poulin2006visual} utilized a simple graphical explanation to interpret the naive Bayesian, linear support vector machine and logistic regression classification process, and provided visualization of the classifier decisions and visualization of the evidence for these decisions. Krause et al. \citep{krause2014infuse} proposed to visualize the ranking information of predictive features to help analysts understand how predictive features are being ranked across feature selection algorithms, cross-validation folds, and classifiers. Visualization can be used to provide an intuitive visual way to understand machine learning algorithms, but it is a better way to understand malware through malicious behaviors. Therefore, in this paper, we try to interpret machine learning algorithms through another way, behavior interpreting.

\subsubsection{Behavior Interpreting}
In order to interpret machine learning models itself, it is crucial to understand how they make predictions, which we define as behavior interpreting here. Through behavior interpreting, we can understand the relation between the input elements and models' output. To achieve this goal, many researchers have tried to combine the elements that have the greatest impact on predictions to explain behaviors.
In 2016, Ribeiro et al. \citep{ribeiro2016should} proposed a model-agnostic method called LIME. It treated the model as a black-box and then generated a linear model to approximate the local part of the model. The authors achieved this purpose by minimizing the expected locally-aware loss. After that, the authors tried to interpret the machine learning result through several features with the most weight. However, because LIME assumes that features are independent, although LIME is designed for explaining the predictions of any classifier, it actually supports CNN to work with image classifiers, but does not well support RNN and MLP. For malware detection, features are interrelated, which makes it difficult for LIME to accurately approximate the decision boundary near an instance.   
In 2018, Guo W et al. \citep{guo2018lemna} proposed LEMNA, a high-fidelity explanation method that solves the problem in LIME. LEMNA utilized fused lasso, which acts as a penalty term that manifests as a constraint imposed upon coefficients in loss functions, to handle the feature dependency problems. Then, it integrated fused lasso into a mixture regression model to more accurately approximate locally nonlinear decision boundaries to support complex deep learning decision. The mixture regression model is a combination of multiple linear regression models. This method also interpreted the model through features with the most weight and is more fidelity than other existing methods. However, there are inevitably deviations due to the use of linear or simple models to approximate the original complex model. Apart from the above work, some survey papers~\citep{lipton2016mythos,guidotti2019survey} also conducted studies on interpretability. {All in all, they cannot interpret models' output accurately in Android malware detection. To solve this problem, we propose an interpretable machine learning model with a customized attention mechanism.}

{\subsection{Applications of Attention Mechanism}}
The attention mechanism is mainly applied to machine translation and computer vision. Bahdanau et al. \citep{bahdanau2014neural} first proposed to solve the problem of incapability of remembering long source sentences in neural machine translation (NMT).
Xu et al. \citep{xu2015show} inspired by the attention mechanism in machine translation, proposed an attention-based model that applied the attention mechanism to images to automatically describe the content of images. They first use a convolutional neural network to extract $L$ feature vectors from the image, each of which is a D-dimensional representation corresponding to a part of the image. Then they use an LSTM decoder to consume the convolution features in order to produce descriptive words one by one, where the weights are learned through attention. The decoder selectively focuses on certain parts of an image by weighting a subset of all the feature vectors. The visualization of the attention weight can indicate the regions of the image that the model pays attention to in order to output a certain word. In addition, it also allows us to understand why some mistakes were made by the model.
Vaswani et al. \citep{vaswani2017attention} proposed a new simple network architecture, the Transformer, based solely on the attention mechanism to perform machine translation tasks, and achieved good performance. There are many other applications for attention mechanism, such as machine reading \citep{cheng2016long}, video summarization \citep{bilkhu2019attention} and document classification \citep{yang2016hierarchical}.
Attention mechanism has been used to accomplish many machine learning tasks and achieved great success. Therefore, we make the first attempt to apply it in malware detection and interpret the classification results, but {the traditional attention mechanism cannot be used directly since its elements and targets are expressed in vector form. We customize the attention mechanism through a fully connected network to learn the correlation between scalar-valued feature elements and assign corresponding weights to the elements. }

\vspace{2mm}
\section{Conclusion}
In this paper, we proposed a novel approach called \mytool to interpret the malicious behaviors of Android apps by leveraging a customized attention mechanism with the MLP model. \mytool achieved a high accuracy in Android malware detection, and output a reasonable natural language description to interpret the malicious behaviors by leveraging the key features pinpointed by the classification phase. Additionally, we compared \mytool with LIME and Drebin, and demonstrated that \mytool obtained better performance in interpretability than the other two methods. 
Finally, we presented an in-depth discussion to highlight the lessons learned and open-challenges in this research field. \revised{The source code is released on the website~{\url{https://github.com/wubozhi/Xmal}}.}

\begin{acks}
This work was supported by Singapore Ministry of Education Academic Research Fund Tier 1 (Award No. 2018-T1-002-069), the National Research Foundation, Prime Ministers Office, Singapore under its National Cybersecurity R\&D Program (Award No. NRF2018 NCR-NCR005-0001), the Singapore National Research Foundation under NCR Award Number NSOE003-0001, NRF Investigatorship NRFI06-2020-0022, {and the
Research Grants Council of the Hong Kong Special Administrative Region, China (No. CUHK 14210717 of the General Research Fund).}
\end{acks}

\bibliographystyle{ACM-Reference-Format}
\bibliography{acmart}

\end{document}